\documentstyle[twoside,axodraw,epsf]{article}

\catcode`\@=11
\long\def\@makefntext#1{
\protect\noindent \hbox to 3.2pt {\hskip-.9pt  
$^{{\eightrm\@thefnmark}}$\hfil}#1\hfill}               

\def\@makefnmark{\hbox to 0pt{$^{\@thefnmark}$\hss}}    
        
\def\ps@myheadings{\let\@mkboth\@gobbletwo
\def\@oddhead{\hbox{}
\rightmark\hfil\eightrm\thepage}   
\def\@oddfoot{}\def\@evenhead{\eightrm\thepage\hfil
\leftmark\hbox{}}\def\@evenfoot{}
\def\sectionmark##1{}\def\subsectionmark##1{}}

\oddsidemargin=\evensidemargin
\newcounter{sectionc}\newcounter{subsectionc}\newcounter{subsubsectionc}
\renewcommand{\section}[1] {\vspace{12pt}\addtocounter{sectionc}{1} 
\setcounter{subsectionc}{0}\setcounter{subsubsectionc}{0}\noindent 
        {\tenbf\thesectionc. #1}\par\vspace{5pt}}
\renewcommand{\subsection}[1] {\vspace{12pt}\addtocounter{subsectionc}{1} 
        \setcounter{subsubsectionc}{0}\noindent 
        {\bf\thesectionc.\thesubsectionc. {\kern1pt \bfit #1}}\par\vspace{5pt}}
\renewcommand{\subsubsection}[1] {\vspace{12pt}\addtocounter{subsubsectionc}{1}
        \noindent{\tenrm\thesectionc.\thesubsectionc.\thesubsubsectionc.
        {\kern1pt \tenit #1}}\par\vspace{5pt}}
\newcommand{\nonumsection}[1] {\vspace{12pt}\noindent{\tenbf #1}
        \par\vspace{5pt}}

\newcounter{appendixc}
\newcounter{subappendixc}[appendixc]
\newcounter{subsubappendixc}[subappendixc]
\renewcommand{\thesubappendixc}{\Alph{appendixc}.\arabic{subappendixc}}
\renewcommand{\thesubsubappendixc}
        {\Alph{appendixc}.\arabic{subappendixc}.\arabic{subsubappendixc}}

\renewcommand{\appendix}[1] {\vspace{12pt}
        \refstepcounter{appendixc}
        \setcounter{figure}{0}
        \setcounter{table}{0}
        \setcounter{lemma}{0}
        \setcounter{theorem}{0}
        \setcounter{corollary}{0}
        \setcounter{definition}{0}
        \setcounter{equation}{0}
        \renewcommand{\thefigure}{\Alph{appendixc}.\arabic{figure}}
        \renewcommand{\thetable}{\Alph{appendixc}.\arabic{table}}
        \renewcommand{\theappendixc}{\Alph{appendixc}}
        \renewcommand{\thelemma}{\Alph{appendixc}.\arabic{lemma}}
        \renewcommand{\thetheorem}{\Alph{appendixc}.\arabic{theorem}}
        \renewcommand{\thedefinition}{\Alph{appendixc}.\arabic{definition}}
        \renewcommand{\thecorollary}{\Alph{appendixc}.\arabic{corollary}}
        \renewcommand{\theequation}{\Alph{appendixc}.\arabic{equation}}
        \noindent{\tenbf Appendix \theappendixc #1}\par\vspace{5pt}}
\newcommand{\subappendix}[1] {\vspace{12pt}
        \refstepcounter{subappendixc}
        \noindent{\bf Appendix \thesubappendixc. {\kern1pt \bfit #1}}
        \par\vspace{5pt}}
\newcommand{\subsubappendix}[1] {\vspace{12pt}
        \refstepcounter{subsubappendixc}
        \noindent{\rm Appendix \thesubsubappendixc. {\kern1pt \tenit #1}}
        \par\vspace{5pt}}

\newcommand{\textlineskip}{\baselineskip=13pt}
\newcommand{\smalllineskip}{\baselineskip=10pt}

\def\eightcirc{
\begin{picture}(0,0)
\put(4.4,1.8){\circle{6.5}}
\end{picture}}
\def\eightcopyright{\eightcirc\kern2.7pt\hbox{\eightrm c}}

\def\abstracts#1#2#3{{
        \centering{\begin{minipage}{4.5in}\baselineskip=10pt\footnotesize
        \parindent=0pt #1\par 
        \parindent=15pt #2\par
        \parindent=15pt #3
        \end{minipage}}\par}} 

\renewenvironment{thebibliography}[1]
        {\frenchspacing
         \ninerm\baselineskip=11pt
         \begin{list}{\arabic{enumi}.}
        {\usecounter{enumi}\setlength{\parsep}{0pt}
         \setlength{\leftmargin 12.7pt}{\rightmargin 0pt} 
         \setlength{\itemsep}{0pt} \settowidth
        {\labelwidth}{#1.}\sloppy}}{\end{list}}

\newcounter{itemlistc}
\newcounter{romanlistc}
\newcounter{alphlistc}
\newcounter{arabiclistc}

\newcommand{\fcaption}[1]{
        \refstepcounter{figure}
        \setbox\@tempboxa = \hbox{\footnotesize Fig.~\thefigure. #1}
        \ifdim \wd\@tempboxa > 5in
           {\begin{center}
        \parbox{5in}{\footnotesize\smalllineskip Fig.~\thefigure. #1}
            \end{center}}
        \else
             {\begin{center}
             {\footnotesize Fig.~\thefigure. #1}
              \end{center}}
        \fi}

\newcommand{\tcaption}[1]{
        \refstepcounter{table}
        \setbox\@tempboxa = \hbox{\footnotesize Table~\thetable. #1}
        \ifdim \wd\@tempboxa > 5in
           {\begin{center}
        \parbox{5in}{\footnotesize\smalllineskip Table~\thetable. #1}
            \end{center}}
        \else
             {\begin{center}
             {\footnotesize Table~\thetable. #1}
              \end{center}}
        \fi}

\def\@citex[#1]#2{\if@filesw\immediate\write\@auxout
        {\string\citation{#2}}\fi
\def\@citea{}\@cite{\@for\@citeb:=#2\do
        {\@citea\def\@citea{,}\@ifundefined
        {b@\@citeb}{{\bf ?}\@warning
        {Citation `\@citeb' on page \thepage \space undefined}}
        {\csname b@\@citeb\endcsname}}}{#1}}

\newif\if@cghi
\def\cite{\@cghitrue\@ifnextchar [{\@tempswatrue
        \@citex}{\@tempswafalse\@citex[]}}
\def\citelow{\@cghifalse\@ifnextchar [{\@tempswatrue
        \@citex}{\@tempswafalse\@citex[]}}
\def\@cite#1#2{{$\null^{#1}$\if@tempswa\typeout
        {IJCGA warning: optional citation argument 
        ignored: `#2'} \fi}}

\def\pmb#1{\setbox0=\hbox{#1}
        \kern-.025em\copy0\kern-\wd0
        \kern.05em\copy0\kern-\wd0
        \kern-.025em\raise.0433em\box0}

\def\fnt#1#2{\footnotetext{\kern-.3em
        {$^{\mbox{\scriptsize #1}}$}{#2}}}

\def\fpage#1{\begingroup
\voffset=.3in
\thispagestyle{empty}\begin{table}[b]\centerline{\footnotesize #1}
        \end{table}\endgroup}

\def\runninghead#1#2{\pagestyle{myheadings}
\markboth{{\protect\footnotesize\it{\quad #1}}\hfill}
{\hfill{\protect\footnotesize\it{#2\quad}}}}
\font\tenrm=cmr10
\font\tenit=cmti10 
\font\tenbf=cmbx10
\font\bfit=cmbxti10 at 10pt
\font\ninerm=cmr9

\font\eightrm=cmr8

\def\qed{\hbox{${\vcenter{\vbox{                        
   \hrule height 0.4pt\hbox{\vrule width 0.4pt height 6pt
   \kern5pt\vrule width 0.4pt}\hrule height 0.4pt}}}$}}

\def\theequation{\arabic{section}.\arabic{equation}}

\begin{document}

\runninghead{Apostolos Pilaftsis}{Heavy Majorana neutrinos and baryogenesis}

\normalsize\textlineskip
\thispagestyle{empty}
\setcounter{page}{1}


\vspace*{-0.5cm}
\begin{flushright}
CERN-TH/98-386\\
hep-ph/9812256\\
December 1998
\end{flushright}
\vspace{-1.cm}

\vspace*{0.88truein}

\fpage{1}
\centerline{\bf HEAVY MAJORANA NEUTRINOS AND BARYOGENESIS
\footnote{To appear in the review section of 
{\em International Journal of Modern Physics A}} }

\vspace*{0.37truein}
\centerline{\footnotesize APOSTOLOS PILAFTSIS\footnote{E-mail address:
pilaftsi@mail.cern.ch}}
\vspace*{0.015truein}
\centerline{\footnotesize\it Theory Division, CERN, CH-1211 Geneva 23,
Switzerland}
\vspace*{0.225truein}

\vspace*{0.21truein} 

\abstracts{The  scenario of    baryogenesis through  leptogenesis   is
  reviewed in models involving heavy  Majorana neutrinos.  The various
  mechanisms of   CP  violation occurring  in   the out-of-equilibrium
  lepton-number-violating   decays of  heavy   Majorana neutrinos  are
  studied within a resummation approach  to unstable-particle mixing.  
  It is explicitly demonstrated how the resummation approach preserves
  crucial field-theoretic   properties   such  as  unitarity   and CPT
  invariance.  Predictions of  representative  scenarios are presented
  after solving numerically  the  Boltzmann equations  describing  the
  thermodynamic  evolution of  the  Universe.   The   phenomenological
  consequences of  loop    effects of  heavy   Majorana neutrinos   on
  low-energy observables, such  as lepton-flavour and/or lepton-number
  non-conservation  in  $\tau$   and  $Z$-boson  decays  and  electron
  electric dipole moment, are discussed.}{}{}


\vspace*{1pt}\textlineskip      
\setcounter{section}{1}
\setcounter{equation}{0}
\section{Introduction}          
\vspace*{-0.5pt}
\noindent
Present astronomical observations related  to abundances of the  light
elements   $^4{\rm He}$  and $^4\overline{\rm  He}$,   the  content of
protons   versus   antiprotons  in cosmic  rays,    etc,  lead  to the
conclusion\cite{RDTMQ,KT}  that, before  the nucleosynthesis epoch, the
Universe must have possessed an excess in  the baryon number $B$ which
is expressed by the small baryon-to-photon ratio of number densities
\begin{equation}
  \label{nB}
\frac{n_{\Delta B}}{n_\gamma}\ =\ (4-7)\times 10^{-10}\ .
\end{equation}
This  baryonic  asymmetry, $n_{\Delta B} =   n_B - n_{\bar{B}} \approx
n_B$, should have survived until today if there  had been no processes
that  violate  the  $B$  number and/or modify   the  number density of
photons $n_\gamma$.  Sakharov,  assuming that the Universe was created
initially in a $B$-conserving symmetric    state, was able to   derive
three necessary conditions\cite{ADS} to  explain the  baryon asymmetry
in the Universe (BAU):
\begin{itemize}
  
\item[  (i)]  There must be  $B$-violating  interactions in nature, so
  that a net $B$ number can in principle be generated.
  
\item[ (ii)] The    $B$-violating interactions   should violate    the
  discrete  symmetries of charge  conjugation  (C) and that  resulting
  from the combined action of  charge and parity transformations (CP). 
  In  this way, an excess in  baryons over antibaryons, $\Delta B$, is
  produced.
  
\item[(iii)] The $B$-   and  CP-violating processes  must  be  out  of
  thermal equilibrium, \cite{KW,EWK&SW}  namely  they  should have  an
  interaction rate smaller than the   expansion rate of the Universe.  
  This last requirement ensures that  the  produced $\Delta B$ is  not
  washed out by the inverse processes.

\end{itemize}

Grand unified theories (GUT's) can  in principle contain all the above
necessary   ingredients for baryogenesis.\cite{MY}  In  such theories,
out-of-equilibrium $B$-  and CP-violating decays of super-heavy bosons
with  masses near to the  grand unification scale $M_X\approx 10^{15}$
GeV can produce the  BAU.  However, this solution  to the BAU has  its
own problems.  The main difficulty is the generic feature that minimal
GUT's predict  very small CP violation, since  it  occurs at very high
orders in  perturbation  theory.   This  problem   may be  avoided  by
augmenting GUT's  with  extra Higgs representations.\cite{RDTMQ} Also,
GUT's must comply with limits obtained by experiments on the stability
of the  proton.  Such experiments put  tight constraints on the masses
of the  GUT bosons mediating $B$  violation and their couplings to the
matter.  Another severe limitation to scenarios of baryogenesis arises
from  the   anomalous  $B+L$-violating    processes, also   known   as
sphalerons,\cite{hooft,sphal,KRS} which are in thermal equilibrium for
temperatures\cite{AMcL,BS}  $200  \stackrel{\displaystyle  <}{\sim}  T
\stackrel{\displaystyle      <}{\sim}  10^{12}$  GeV.    Unlike $B+L$,
sphalerons   preserve  the quantum    number  $B-L$.  Therefore,   any
primordial  BAU  generated  at  the   GUT scale  should   not rely  on
$B+L$-violating operators,  which   imposes   a   further  non-trivial
constraint on unified  theories.   In that vein, Kuzmin,  Rubakov  and
Shaposhnikov\cite{KRS}   suggested     that    the same      anomalous
$B+L$-violating  electroweak  interactions  may  produce  the observed
excess in $B$ during  a first-order electroweak phase transition. Such
a   mechanism  crucially    depends     on  the   Higgs-boson     mass
$M_H$,\cite{ADD,RS} and the experimental fact $M_H>80$ GeV practically
rules out   this scenario  of electroweak  baryogenesis.\cite{EWphase}
Therefore, baryogenesis  provides the strongest indication against the
completeness of  the  SM, as  well  as  poses limits  on its  possible
new-physics extensions.

Among  the many baryogenesis scenarios  invoked in the literature, the
most attractive one is due to  Fukugita and Yanagida\cite{FY}, and our
emphasis will be put  on their  scenario  in this review  article.  In
such   a scenario, out-of-equilibrium   $L$-violating decays  of heavy
Majorana neutrinos $N_i$, with  masses  $m_{N_i} \gg T_c$, produce  an
excess in the lepton  number $L$ which  is converted into the  desired
excess  in $B$  by  means  of $B+L$-violating sphaleron  interactions,
which are in thermal equilibrium above the critical temperature $T_c$.
Over the last years, many authors have discussed such a scenario, also
known                as                baryogenesis            through
leptogenesis.\cite{MAL,CEV,epsilonprime,LS1,Paschos,APRD}  However, we
should  remark that isosinglet   heavy neutrinos are not theoretically
compelling in order to create an excess in  the $L$ number.  Recently,
Ma and Sarkar\cite{Ma/Sarkar} suggested  a leptogenesis scenario based
on    a generalized Higgs-triplet model,\cite{Triplet1,Triplet2} where
the leptonic asymmetry is generated by out-of-equilibrium CP-violating
decays  of heavy doubly charged Higgs  triplets  into charged leptons. 
However, such alternatives seem to face the known gravitino problem if
they are to be embedded in  a supersymmetric theory.\cite{Del/Sar} The
charged Higgs  triplets or their  supersymmetric partners may interact
strongly  with gravitinos and  produce them in  large abundances.  The
slow  decay  rate  of gravitinos   during   the nucleosynthesis  epoch
distorts the abundances of the light  elements at a level inconsistent
with present observations.

Mechanisms  that   enhance  CP violation    play   a decisive  role in
baryogenesis.  Using the  terminology  known from  the  $K^0\bar{K}^0$
system,\cite{reviewCP} one may distinguish the following two cases:
\begin{itemize}
  
\item[ (i)] CP violation originating from the interference between the
  tree-level  decay amplitude and the  absorptive part of the one-loop
  vertex. Such a   mechanism is usually called  $\varepsilon'$-type CP
  violation.\cite{FY,MAL,CEV,epsilonprime}
  
\item[(ii)] CP violation induced by the interference of the tree-level
  graph and the absorptive part of  a one-loop self-energy transition. 
  This mechanism is termed  $\varepsilon$-type CP
  violation.\cite{IKS,BR,KRS,LS1,Paschos,APRD}

\end{itemize}

\begin{figure}
\begin{center}
\begin{picture}(360,110)(0,0)
\SetWidth{0.8}

\ArrowLine(0,70)(30,70)\ArrowLine(30,70)(70,70)
\Line(70,70)(100,70)\DashArrowArc(50,70)(20,0,180){5}
\Text(85,70)[]{{\boldmath $\times$}}
\Text(0,77)[bl]{$N_i$}\Text(85,77)[b]{$N_j$}
\Text(50,95)[b]{$\Phi^\dagger$}\Text(50,65)[t]{$L$}
\ArrowLine(130,40)(100,70)\DashArrowLine(100,70)(130,90){5}
\Text(140,30)[r]{$L^C$}\Text(140,100)[r]{$\Phi$}
\Text(80,10)[]{\bf (a)}

\ArrowLine(250,70)(280,70)\ArrowLine(280,70)(310,70)\Line(310,70)(310,30)
\ArrowLine(340,30)(310,30)\DashArrowLine(310,70)(340,90){5}
\DashArrowLine(310,30)(280,70){5}\Text(310,50)[]{{\boldmath $\times$}}
\Text(250,65)[lt]{$N_i$}\Text(295,80)[]{$L$}\Text(315,50)[l]{$N_j$}
\Text(345,90)[l]{$\Phi$}\Text(345,30)[l]{$L^C$}\Text(290,50)[r]{$\Phi^\dagger$}
\Text(280,10)[]{\bf (b)}

\end{picture}\\[0.4cm]
\end{center}

\fcaption{One-loop (a) self-energy and (b) vertex graphs in heavy Majorana
  neutrino decays.}\label{fig:0}

\end{figure}

As can be  seen   from Fig.\ \ref{fig:0},    both  of the above    two
mechanisms of  CP violation are present\cite{LS1,Paschos,APRD}  in the
usual  leptogenesis scenario\cite{FY}    of  heavy Majorana   neutrino
decays.    CP violation of   the  $\varepsilon'$ type was  extensively
discussed in  the  literature.  \cite{FY,MAL,CEV,epsilonprime}  If all
Yukawa couplings of the Higgs fields to $N_i$  and the ordinary lepton
isodoublets are    of  comparable  order,\cite{MAL,epsilonprime}  then
baryogenesis through the  $\varepsilon'$-type mechanism requires  very
heavy Majorana neutrinos with masses of order $10^7$--$10^8$ GeV. Such
a high  mass bound may  be  lifted if a  strong  hierarchy for  Yukawa
couplings and $m_{N_i}$ is assumed.\cite{MAL,CEV} However, without the
latter       assumption,\cite{epsilonprime}            one     obtains
$\varepsilon'<10^{-15}$  for $m_{N_i}\approx   1$ TeV,  and hence very
heavy neutrinos are needed to account for the BAU.

Recently, $\varepsilon$-type CP violation and its implications for the
BAU has  received much attention.\cite{Paschos,APRD} In particular, it
has  been   observed\cite{Paschos,APRD}  that   CP  violation can   be
considerably enhanced  through the   mixing of two   nearly degenerate
heavy Majorana neutrinos. Such an  analysis cannot be performed in the
conventional     field-theoretic    framework,  since     finite-order
perturbation theory breaks down in the  limit of degenerate particles. 
To   be  specific,  the  wave-function  amplitude   that describes the
CP-asymmetric mixing of two heavy Majorana neutrinos, $N_1$ and $N_2$,
say,   is    inversely proportional      to    the  mass     splitting
$m_{N_1}-m_{N_2}$,  and  it becomes singular if   degeneracy is exact. 
Solutions to this problem have been based  on the Weisskopf and Wigner
(WW)\cite{WW}   approximation,\cite{Paschos} and  the      resummation
approach.\cite{ANPB,APRD} Both approaches  lead to similar conclusions
concerning  the resonant enhancement of CP  violation.  Here, we shall
follow   the   latter method, as    the   discussion  of many  crucial
field-theoretic issues,  such as  renormalization, CPT  invariance and
unitarity, is conceptually more intuitive in this framework.

To describe the  dynamics  of  CP  violation  through  mixing of   two
unstable  particles, one  is    compelled   to rely  on    resummation
approaches,  which   treat   unstable  particles    in a    consistent
way.\cite{APCP,unstable} In fact,  to any finite order in perturbation
theory, physical amplitudes  reflect the local gauge symmetry, respect
unitarity, are invariant under  the renormalization group, and satisfy
the equivalence theorem.  All  of the above  properties should also be
present after  resummation.  Unfortunately, resummation  methods often
end up   violating one or  more of   them. The  reason is that  subtle
cancellations  are distorted when  certain parts of  the amplitude are
resummed   to  all  orders  in  perturbation  theory, whereas  others,
carrying  important  physical information, are   only  considered to a
finite order.   In this  context,   a novel  diagrammatic  resummation
approach has  been  developed,\cite{PP} which is   based on  the pinch
technique (PT)\cite{Pinch} and  devoid of the  above pathologies.   In
the PT resummation approach,  basic field-theoretic requirements, such
as    analyticity,    unitarity,        gauge    invariance        and
renormalizability,\cite{PP}  are naturally  satisfied.  Apart from the
great phenomenological importance of such  a resummation formalism for
the proper definition of the mass and the width of unstable particles,
such as the $W$,  the $Z$ boson and  the Higgs boson,\cite{PP,ET} this
formalism may also  be extended  to the  case  of mixing between   two
intermediate   resonant states in scattering processes\cite{APRL,ANPB}
retaining all the required field-theoretic properties mentioned above.

The afore-mentioned resummation formalism has  been proved to be  very
successful in describing resonant transitions taking place in collider
experiments.  These  are situations where  the  unstable particles are
produced   by  given  asymptotic  states,  $e^+e^-$,   say,  and their
subsequent decay is observed by detecting some other asymptotic states
in the final state, e.g.\  $e^+e^-\to Z^*\to \mu^+\mu^-$.  However, in
an  expanding   Universe, the unstable  particles may   undergo a huge
number  of collisions  before they  eventually  decay.   Each of these
collisions  contributes a Coulomb  phase  shift, and  hence the  mixed
heavy particles are practically uncorrelated when they decay.  To some
extent, this thermodynamic phenomenon  may  be described by  Boltzmann
equations.\cite{KW}  In this context,  a  related formalism for decays
has been  developed,\cite{APRD}  which effectively  takes into account
{\em decoherence} phenomena in  the   mixing and subsequent  decay  of
heavy  particles,  namely heavy   Majorana   neutrinos  in our  case.  
Specifically, it is  shown  that $\varepsilon$-type  CP  violation can
even  be of order unity.\cite{APRD}  This is in agreement with earlier
studies on resonant   CP   violation through mixing   in   scatterings
involving top quarks, supersymmetric quarks or  Higgs particles in the
intermediate state.\cite{APCP,APRL,ANPB} Finally,  we must remark that
alternative formulations of  Boltzmann equations already  exist in the
recent    literature\cite{BEqs} but  they are expected    not to alter
drastically the  existing conclusions\cite{Paschos,APRD} as far as the
resonant phenomenon of CP violation is concerned.

The organization of the review article is as  follows: in Section 2 we
briefly  review the basic theoretical  background concerning the $B+L$
anomaly in   the Standard  Model (SM), and   the  effect  of sphaleron
processes on the   chemical    potentials of SM   particles.     These
considerations  lead to  a  relation  between  the generated  leptonic
asymmetry and the   observed baryonic asymmetry  induced  by sphaleron
interactions.  In Section 3 we discuss theories that naturally include
heavy Majorana neutrinos.   For illustration,  we  consider a  minimal
model  with two isosinglet neutrinos and  demonstrate how CP and L can
simultaneously be violated in this model.  In Section 4 we address the
issue of renormalizability of the  minimal iso-singlet neutrino model. 
Section 5  discusses in   detail  the  resummation approach  and   its
effective extension to  describe  incoherent decays of heavy  unstable
fermions.  In Section 6 we apply the  effective approach to the decays
of heavy  Majorana neutrinos.  In Section  7 we explicitly demonstrate
how the resummation  approach satisfies unitarity.   In Section 8,  we
solve   numerically    the  Boltzmann  equations  for   representative
leptogenesis scenarios, and  give numerical  estimates and comparisons
for     the     BAU  generated      through     $\varepsilon$-  and/or
$\varepsilon'$-type CP violation.  Furthermore, we estimate the impact
of    finite-temperature  effects on  the  resonant   phenomenon of CP
violation.    Heavy Majorana neutrinos     may  also  have   important
phenomenological implications  for low-energy observables, as they can
give rise  to  a non-vanishing  electric  dipole  moment (EDM) of  the
electron at two loops or induce  $L$-violating decays of the $Z$ boson
and  the $\tau$ lepton.    These new-physics effects  are  detailed in
Section 9.  Section 10 summarizes our conclusions.


\setcounter{section}{2}
\setcounter{equation}{0}
\section{$B+L$ anomaly and sphaleron processes \label{sec:2}}
\noindent
In the SM, the $B$ and $L$ numbers are only conserved in the classical
action.   After  quantization,  however, both   baryonic  and leptonic
currents are violated by triangle anomalies, i.e.,
\begin{equation}
  \label{anomaly}
\partial_\mu J^\mu_B\ =\ \partial_\mu J^\mu_L\ =\ i\,
\frac{N_F}{8\pi}\ \Big( -\alpha_w W^{\mu\nu,a}\widetilde{W}^a_{\mu\nu}
+ \alpha_{{}_Y} Y^{\mu\nu}\widetilde{Y}_{\mu\nu}\, \Big)\, , 
\end{equation}
where $N_F$ is the number of  flavours, and $\alpha_w = g^2_w/(4\pi)$,
$\alpha_{{}_Y} =  g^2_{{}_Y}/(4\pi)$,  are the SU(2)$_L$  and U(1)$_Y$
fine-structure constants, respectively.     Similarly,   $W^{\mu\nu}$,
$Y^{\mu\nu}$ are  their respective   field-strength tensors, and   the
antisymmetric  tensors    $\widetilde{W}_{\mu\nu}      =   \frac{1}{2}
\varepsilon_{\mu\nu\lambda\rho}                      W^{\lambda\rho}$,
$\widetilde{Y}_{\mu\nu} =  \frac{1}{2} \varepsilon_{\mu\nu\lambda\rho}
Y^{\lambda\rho}$ are  their  associated duals.  Furthermore,  baryonic
and leptonic currents are defined as
\begin{eqnarray}
  \label{JB}
J^\mu_B &=& \frac{1}{3}\ \sum_{q,\alpha} \bar{q}^\alpha \gamma^\mu
q^\alpha\, \\
  \label{JL}
J^\mu_L &=& \sum_{l,\nu_l}\, (\, \bar{l} \gamma^\mu
l\, +\ \bar{\nu}_l \gamma^\mu \nu_l\, )\, ,
\end{eqnarray}
where   $q$,  $l$ and $\nu_l$   denote   quarks, charged  leptons  and
neutrinos,  respectively, and the  index $\alpha$ indicates the colour
degrees of freedom of the   quarks.  Since Eq.\ (\ref{anomaly})   also
holds for individual  lepton families, the actual anomaly-free charges
are
\begin{equation}
  \label{Li}
\frac{1}{3} B\ -\ L_e\, ,\quad \frac{1}{3} B\ -\ L_\mu\, ,\quad
\frac{1}{3} B\ -\ L_\tau\, . 
\end{equation}
It is then  obvious that $B+L$  symmetry is anomalously  broken at the
quantum level.

The   different gauge  field    configurations  are  characterized  by
different Chern-Simons numbers $n_{\rm CS}$.  The CS numbers label the
infinitely many degenerate vacua of the system. The variation of $B+L$
number due to a quantum tunnelling from  one vacuum state into another
is given by
\begin{equation}
  \label{DBL}
\Delta (B+L) \ =\ 2N_F\, \frac{\alpha_w}{8\pi}\int d^4x\ 
W^{\mu\nu,a}\widetilde{W}^a_{\mu\nu}\ =\ 2N_F \Delta n_{\rm CS}\, .
\end{equation}
At zero temperature, 't Hooft\cite{hooft} estimated the probability of
$B$-violating processes, and found  them to be extremely suppressed by
a  factor $\exp (-4\pi n_{\rm CS}/\alpha_w)  \approx \exp (-150 n_{\rm
  CS})$ relative to the $B$-conserving ones with $n_{\rm CS} = 0$.

The situation changes drastically at  finite temperatures.  The effect
of    non-trivial  topological   instanton-type   solutions,    termed
sphalerons,\cite{sphal}   is  amplified at high  temperatures, thereby
enhancing also the rate of the $B$-violating processes. To be precise,
sphaleron interactions are in thermal  equilibrium for temperatures in
the interval
\begin{equation}
  \label{Tsphal}
100\ {\rm GeV}\ \stackrel{\displaystyle <}{\sim }\ T\
\stackrel{\displaystyle <}{\sim }\ 10^{12}\ {\rm GeV}\, . 
\end{equation}
Sphalerons may be  thought of as the creation  out of the vacuum  of a
state
\begin{equation}
  \label{vsphal}
\Pi_{i=1,N_F}\ (u_L d_L d_L \nu_L)_i\ .
\end{equation}
Since  these  interactions   violate $B+L$,  any   primordial baryonic
asymmetry  $B$ should have a significant  component in $B-L$ or in the
charges stated in Eq.\ (\ref{Li}),  which are preserved by sphalerons,
whereas any   $B+L$ component will be   washed  out.  Decays  of heavy
Majorana  neutrinos produce  an excess in  $L$, which  can  naively be
written  as  a  sum of  an  excess   in  $\frac{1}{2} (B+L)$   and  in
$\frac{1}{2} (B-L)$.  Sphalerons  will then erase  the $B+L$ component
but preserve the $B-L$, so one expects that about half of the leptonic
asymmetry  $L$ will be converted  into the baryonic asymmetry $B$, and
also be preserved  as $B-L$ asymmetry.  As  we will see below,  a more
careful analysis based on  chemical potentials leads to the conclusion
that sphalerons approximately convert one-third of an initial leptonic
asymmetry $L$ into the observed baryonic asymmetry $B$.

For  illustration, we shall  assume that  all SM  particles are almost
massless  at  temperatures  above the    critical  temperature $T_c$.  
Actually, they have  thermal masses but, to  leading order,  these are
small and may be neglected. The  number density of a particle $\alpha$
is given by
\begin{equation}
  \label{nalpha}
n_\alpha\ =\ g_\alpha\, \int\, \frac{d^3\vec{p}_\alpha}{2E_\alpha
(2\pi)^3}\
\frac{1}{\exp [ (E_\alpha  - \mu_\alpha )/T ]\, \pm 1 }\, ,
\end{equation}
where $g_\alpha$  counts the internal degrees  of freedom of $\alpha$,
$\vec{p}_\alpha$      and   $E_\alpha     =   (|\vec{p}_\alpha|^2    +
m_\alpha)^{1/2}$ are the three-momenta and the energy of the particle,
respectively.  The plus sign  in Eq.\ (\ref{nalpha}) is  for particles
obeying the   Fermi-Dirac  statistics  and  the  minus  for  particles
governed by the Bose-Einstein statistics.   The chemical potential for
anti-particles, e.g.\ that of  $\bar{\alpha}$, is opposite to that  of
the particles,  i.e.\  $\mu_\alpha = -\mu_{\bar{\alpha}}$.   The later
relation is   valid if particles  and  antiparticles  have interaction
rates   with photons or  other gauge  particles   much higher than the
expansion rate  of the Universe.   This is almost  the case for all SM
particles.  However, this is  not generally true for non-SM particles,
such as  isosinglet or right-handed neutrinos, which   do not have any
tree-level coupling  to the $W$- and  $Z$- bosons; their couplings are
suppressed    by loops  and    small Yukawa   couplings.  Under  these
assumptions,  the number-density asymmetry  of  a SM particle $\alpha$
versus its antiparticle $\bar{\alpha}$ is easily estimated by
\begin{equation}
  \label{nasym}
n_{\Delta \alpha}\ =\ n_\alpha\, -\, n_{\bar{\alpha}}\ \approx\
\frac{g_\alpha}{\pi^2}\ T^3\ \Big(\frac{\mu_\alpha}{T}\Big)\ .  
\end{equation}

We shall now  turn to an analysis of  chemical  potentials in the  SM. 
Since FCNC interactions  are   sufficiently fast, we assign   the same
chemical  potential for all different families  of up and down quarks,
i.e.\ $(\mu_{u_L} , \mu_{d_L},  \mu_{u_R} , \mu_{d_R})$.  In  contrast
to quarks,  individual leptons possess  different chemical potentials,
i.e.\  $(\mu_{l_L},  \mu_{\nu_{lL}}, \mu_{l_R})$, where   $l = e, \mu,
\tau$.   Furthermore,  the chemical  potential   of all neutral  gauge
bosons, such  as gluons, photons, and $Z$  bosons, vanish, and $\mu_W$
is the chemical potential of the $W^-$ boson.  Finally, the components
of the Higgs  doublet  $[\chi^-, \phi^0 =  (H-i\chi^0)/\sqrt{2}]$ have
chemical potentials $(\mu_0, \mu_-)$.  Many chemical potentials can be
eliminated by means of chemical equilibrium  reactions in the SM. More
explicitly, we have
\begin{equation}
  \label{chem}
\begin{array}{rclcrcl}
W^- &\leftrightarrow&\bar{u}_L + d_L\, ,& ~~~~ &  \mu_W &=& -\mu_{u_L} +
\mu_{d_L},\\
W^- &\leftrightarrow& \bar{\nu}_{lL} + l_L\, ,& &  
\mu_W & =& -\mu_{\nu_{lL}} + \mu_{l_L}, \\
W^- &\leftrightarrow& \chi^- + \phi^0\, ,&& \mu_W &=& \mu_- +
\mu_0, \\
\phi^0 &\leftrightarrow& \bar{u}_L + u_R\, , && \mu_0 &=& -\mu_{u_L} +
\mu_{u_R}, \\
\phi^0 &\leftrightarrow& \bar{d}_L + d_R\, , && \mu_0 &=& -\mu_{d_L} +
\mu_{d_R}, \\
\phi^0 &\leftrightarrow& \bar{l}_L + l_R\, , && 
\mu_0 &=& -\mu_{l_L} + \mu_{l_R}.
\end{array}
\end{equation}
As independent  parameters,  we consider $\mu_u  = \mu_{u_L}$,  $\mu =
\sum_l \mu_{\nu_{lL}} = \sum_l   \mu_{l_L}$, $\mu_0$ and  $\mu_W$.  In
the  SM with $N_F$  families and $N_H$  Higgs doublets, the baryon and
lepton number  $B$  and $L$  as well  as  the electric charge  $Q$ and
hypercharge  $Q_3$ may be expressed in   terms of these quantities, as
follows:
\begin{eqnarray}
  \label{BLQQ}
B &=& 4N_F\mu_u\, +\, 2N_F\mu_W\, , \nonumber\\
L &=& 3\mu\, +\, 2N_F\mu_W\, -\, N_F \mu_0\, ,\nonumber\\
Q &=& 2N_F\mu_u\, -\, 2\mu\, +\, 2(2N_F + N_H)\mu_0 
- 2(2N_F + 2 + N_H) \mu_W\, ,\nonumber\\
Q_3 &=& - (2N_F + 4 + 2N_H) \mu_W\, .
\end{eqnarray}
Furthermore, the sphaleron   interactions in Eq.\  (\ref{vsphal}) give
rise to the additional relation
\begin{equation}
  \label{chemspal}
N_F(3\mu_u\ +\ 2\mu_W)\ +\ \mu\ =\ 0\, . 
\end{equation}
Above the electroweak phase transition, both charges $Q$ and $Q_3$ are
conserved, i.e.\ $\langle Q \rangle =  \langle Q_3 \rangle = 0$. Thus,
we  have: $\mu_W   =    0$, $\mu =    -3N_F\mu_u$,  and  $\mu_0 =    -
8N_F\mu_u/(4N_F + 2N_H)$. Using these relationships among the chemical
potentials, it is not difficult to obtain\cite{HT,Dreiner/Ross}
\begin{equation}
  \label{BLrel}
B(T > T_c) \ =\ \frac{8N_F + 4N_H}{22N_F + 13 N_H}\ (B - L)\ .  
\end{equation}
{}From Eq.\ (\ref{BLrel}),  one concludes that almost independently of
the number of generations and Higgs doublets, roughly one-third of the
initial  $B-L$ and/or $L$ asymmetry  will be reprocessed by sphalerons
into an  asymmetry in $B$. This amount  of $B$ asymmetry persists even
after the electroweak phase transition.

\setcounter{section}{3}
\setcounter{equation}{0}
\section{Models with heavy Majorana neutrinos}\label{sec:3}
\noindent
GUT's such as SO(10) \cite{FM,Wol/Wyl} or E${}_6$ \cite{witten} models
naturally  predict  heavy  Majorana  neutrinos.   These  theories also
contain several other particles, e.g.\ leptoquarks, additional charged
and   neutral gauge   bosons,   etc.,  which   may  have   significant
interactions with  heavy Majorana neutrinos and   so affect the number
density of heavy neutrinos. To  avoid excessive complication, we shall
assume that  these  new particles are  much heavier  than the lightest
heavy Majorana  neutrino,  and   are therefore expected   to  decouple
sufficiently fast from the process of leptogenesis.

As   already    mentioned,  SO(10)  \cite{FM,Wol/Wyl}  and/or  E${}_6$
\cite{witten}   models  may   naturally  accommodate   heavy  Majorana
neutrinos. Specifically, SO(10) models can break  down to the SM gauge
group in the following schematic way:
\begin{eqnarray}
\mbox{SO(10)} &\to & G_{422}=\mbox{SU}(4)_{\mbox{\scriptsize PS}}
                              \otimes\mbox{SU}(2)_R \otimes 
                                                     \mbox{SU}(2)_L\nonumber\\
&\to & G_{3221} =
\mbox{SU}(3)_c\otimes \mbox{SU}(2)_R\otimes \mbox{SU}(2)_L\otimes 
                                          \mbox{U}(1)_{(B-L)}\nonumber\\
&\to & \mbox{SM} = G_{321} = \mbox{SU}(3)\otimes \mbox{SU}(2)_L\otimes 
\mbox{U}(1)_Y \, , 
\end{eqnarray}
where  the    subscript  PS    characterizes  the    Pati-Salam  gauge
group.\cite{PS} The spinor representation of  SO(10) is 16-dimensional
and its decomposition under $G_{422}$ reads
\begin{equation}
\label{G422}
G_{422}\, :\ \mbox{{\bf 16}}\ \to\ (4,1,2)\, \oplus\, (\bar{4},2,1)\, .
\end{equation}
Evidently,  SO(10)  contains   the  left-right-symmetric  gauge  group
SU(2)$_R\otimes$SU(2)$_L\otimes$  U(1)$_{(B-L)}$,   which necessitates
the presence of right-handed neutral  leptons. In this scenario, there
can  exist   several Higgs-boson  representations    that may cause  a
breaking of  the groups $G_{422}$ and  $G_{3221}$ down to the SM gauge
group $G_{321}$.\cite{Wol/Wyl,PalMoh}

E$_6$ theories \cite{witten} may  also have a breaking pattern related
to SO(10) theories.  In   fact,  the {\bf 27} spinor    representation
decomposes into  {\bf 16} $\oplus$  {\bf  10} $\oplus$ {\bf  1}  under
SO(10).  This  leads   to four singlet neutrinos   per  SM family: one
neutrino as isodoublet member in {\bf 16}, two neutrinos as isodoublet
members in {\bf 10}, and  one singlet neutrino  in  {\bf 1}. In  these
models,   it is  argued that  depending   on the representation of the
E${}_6$ Higgs multiplets,   two  of  the  four isosinglets  can   have
Majorana masses of few TeV,\cite{witten}  whereas the other two may be
very heavy with masses of the order of the unification scale.

We shall   now discuss a generic  subgroup  that  may be  derived from
SO(10) and/or E$_6$ models.  This generic  subgroup may be realized in
the usual SM  augmented  by a  number $n_R$  of right-handed neutrinos
$\nu_{Ri}$, with $i=1, 2, \dots, n_R$.  As we have discussed above, in
E$_6$ theories the active isosinglet neutrinos may be more than three.
In SO(10) models, left-right symmetry is  more naturally realized with
one right-handed neutrino  per family  (for interesting  alternatives,
see Ref.\cite{Wol/Wyl}). For the sake of generality, we shall keep the
number of iso-singlet neutrinos arbitrary. For definiteness, the quark
sector of the minimal model has the SM form, while the leptonic sector
consists of the fields:
\begin{displaymath}
\left( \begin{array}{c} \nu_{lL} \\ l_L \end{array} \right)\ ,\qquad l_R\ ,
\qquad \nu_{Ri}\ ,
\end{displaymath}
where $l=e,\mu,\tau$.   At  $T\gg T_c \stackrel{\displaystyle >}{\sim}
v$, the vacuum expectation value (VEV) of  the SM Higgs doublet $\Phi$
at temperature  $T$ (with $v=v(0)$) vanishes,  $v(T)=0$. At these high
temperatures, all  SM  particles including Higgs  fields are massless;
they only acquire thermal masses. However, one may have Majorana
masses in the Lagrangian given by
\begin{equation}
\label{Majmass}
-{\cal L}_M\ =\ \frac{1}{2}\, \sum\limits_{i,j=1}^{n_R}\, 
\Big( \bar{\nu}^C_{Ri} M^\nu_{ij} \nu_{Rj}\ +\ 
\bar{\nu}_{Ri} M^{\nu *}_{ij} \nu^C_{Rj}\, \Big)\, .
\end{equation}
where $M^\nu$   is an  $n_R\times  n_R$-dimensional  symmetric matrix,
which is in general complex.  In Eq.\ (\ref{Majmass}), the superscript
$C$ denotes  the operation of   charge conjugation, which acts  on the
four-component chiral spinors $\psi_L$ and $\psi_R$ as follows:
\begin{equation}
  \label{psiC}
(\psi_L)^C\ =\ P_R C\bar{\psi}^T\, , 
\qquad (\psi_R)^C\ =\ P_L C\bar{\psi}^T\, ,  
\end{equation}
where  $P_{L(R)}=[1  -(+)\gamma_5]/2$   is  the chirality   projection
operator. The  mass matrix $M^\nu$ can  be diagonalized by means  of a
unitary transformation
\begin{equation}
  \label{diagMnu}
U^T M^\nu U\ =\ \widehat{M}^\nu\, ,  
\end{equation}
where  $U$ is an  $n_R\times n_R$-dimensional  unitary  matrix and the
diagonal matrix $\widehat{M}^\nu$   contains the $n_R$ heavy  Majorana
masses. Then, the respective $n_R$  mass eigenstates $N_i$ are related
to the flavour states $\nu_{Ri}$ through
\begin{equation}
  \label{nuRi}
\nu_{Ri}\ =\ P_R\sum_{j=1}^{n_R} U_{ij} N_j\, ,\qquad
\nu^C_{Ri}\ =\ P_L\sum_{j=1}^{n_R} U^*_{ij} N_j\, .
\end{equation}

In  the mass  basis  of heavy  Majorana neutrinos,   the Yukawa sector
governing the interactions  of the heavy neutrinos  with the Higgs and
lepton isodoublets is given by
\begin{equation}
\label{LYint}
{\cal L}_Y\ =\ -\, \sum\limits_{l=1}^{n_L} \sum\limits_{j=1}^{n_R}\,
h_{lj}\, (\bar{\nu}_{lL},\ \bar{l}_L)\, 
\left( \begin{array}{c} (H\, -\, i\chi^0)/\sqrt{2} \\ -\, \chi^- \end{array}
\right)\, N_j\ +\ \mbox{H.c.}
\end{equation}
At high  temperatures, the CP-even Higgs  field $H$, the  CP-odd Higgs
scalar $\chi^0$ and the  charged   Higgs scalars $\chi^\pm$  given  in
(\ref{LYint})  are  massless.  In the low-$T$   limit  $T\ll T_c$, the
field $H$   becomes the massive SM  Higgs  boson, whereas $\chi^0$ and
$\chi^\pm$ are the would-be Goldstone bosons eaten by the longitudinal
degrees of freedom of the gauge bosons $Z$  and $W^\pm$, respectively. 
In   the  calculations in Section    6,   we shall include the   $M_H$
dependence in the CP asymmetries.

Let  us now consider a simple  one-generation model where the standard
fermionic content is   extended  by adding  two isosinglet  neutrinos,
e.g.\ $\nu_R$  and $S_L$.  Then, the most   general Yukawa sector that
preserves lepton number reads
\begin{equation}
\label{WW}
- {\cal L}\ =\ \frac{1}{2}\, (\bar{S}_L,\ (\bar{\nu}_R)^C )\, 
\left( \begin{array}{cc}
0 & M \\
M & 0 \end{array} \right)\, \left( \begin{array}{c}
(S_L)^C \\ \nu_R \end{array} \right)\ +\ 
h_R\, (\bar{\nu}_L,\ \bar{l}_L) \tilde{\Phi} \nu_R\ +\ \mbox{H.c.},
\end{equation}
where  $\tilde{\Phi}=i\sigma_2\Phi$   is the isospin  conjugate  Higgs
doublet  and  $\sigma_2$ is the    usual Pauli matrix.  The  kinematic
parameters $M$ and $h_R$  may in general  be complex but their  phases
are not physical. One can make both  real by appropriate redefinitions
of the fermionic fields, i.e.
\begin{equation}
\label{CProt}
L^T_L \equiv (\nu_L, l_L) \to e^{i\phi_L} L^T_L \, ,\qquad 
\nu_R \to e^{i\phi_R} \nu_R,\qquad
S_L \to e^{i\phi_S} S_L .
\end{equation}
One choice  could  be: $\phi_R  = 0$,  $\phi_S  = {\rm  arg}(M)$,  and
$\phi_L = {\rm arg}(h_R)$.  Retaining  the $L$-conserving structure of
the   isosinglet mass matrix, such   scenarios  require a  non-trivial
mixing among   the  generations to  describe CP   violation.\cite{BRV}
Furthermore,  such  scenarios do  not  produce  the necessary leptonic
asymmetry     for baryogenesis; however,   see   Ref.\cite{ARS} for an
interesting variant based on individual lepton-flavour violation.

In order to break both $L$ and CP symmetries of the Lagrangian in Eq.\ 
(\ref{WW}), one must consider at least two extensions in the model:
\begin{itemize}
  
\item[ (i)] The inclusion of two complex $L$-violating Majorana masses
  $\mu_R\bar{\nu}_R\nu^C_R$ and $\mu_L\bar{S}^C_L S_L$.
  
\item[(ii)] The addition   of   the   $L$-violating coupling    $h_L\,
  (\bar{\nu}_L,\     \bar{l}_L) \tilde{\Phi}       (S_L)^C$    and one
  $L$-violating mass parameter, e.g.\ $\mu_R\bar{\nu}_R\nu^C_R$.

\end{itemize}
The two models are  related by a   unitary rotation and  are therefore
equivalent.   The necessary conditions  for CP invariance in these two
scenarios may be found to be
\begin{eqnarray}
\label{CPi_ii}
\mbox{(i)}  && |h_R|^2\, {\rm Im} ( M^{*2}\mu_L\mu_R )\ =\ 0\, ,\nonumber\\
\mbox{(ii)} &&  {\rm Im} (h_Lh_R^*\mu_R M^*)\ =\ 0\, . 
\end{eqnarray}

It is now  interesting to remark that $\mu_L$  and $\mu_R$ can be much
smaller than $M$ within E$_6$ scenarios.\cite{witten} These parameters
may be  induced  after  integrating   out high-dimensional   operators
involving  ultra-heavy non-active neutrinos.   One  may think that the
lepton number is  somehow violated, at the  GUT or Planck scale $M_X$,
by  these additional     non-active  isosinglet fields,   and   it  is
communicated to   the active isosinglet  sector  where $M\ll M_X$.  In
this  way, one can  naturally obtain  a  see-saw-like relation for the
sizes of $\mu_L$ and $\mu_R$, i.e.
\begin{equation}
  \label{muLR}
\mu_L,\ \mu_R\ \sim\ \frac{M^2}{M_X}\quad {\rm or}\quad
\frac{M^2}{M_S}\ ,  
\end{equation}
where $M_S\approx  10^{-3}\, M_X$ could  be some  intermediate see-saw
scale. In such generic mass models, the heavy Majorana neutrinos $N_1$
and $N_2$  have a very  small  mass splitting given by 
\begin{equation}
  \label{xN}
x_N\ =\ \frac{m_{N_2}}{m_{N_1}}\ -\ 1\ \sim\  \frac{\mu_L}{M}\quad
{\rm or}\quad \frac{\mu_R}{M}\ .  
\end{equation}
For  instance, if  $M=10$ TeV  and  $\mu_L=\mu_R = M^2/M_X$, one  then
finds $x_N\approx 10^{-12}$--$10^{-11}$.  As we will  see in Section 6,
such small  values of $x_N$  can lead to  a resonant enhancement of CP
asymmetries in the heavy Majorana neutrino decays.

To obtain the sufficient and necessary conditions of CP invariance for
any flavour structure of  the one-generation model with two isosinglet
neutrinos,  one should   use    a more general  approach,  based    on
generalized CP transformations \cite{BBG} for the fermionic fields:
\begin{equation}
  \label{genCP}
L_L \to e^{i\phi_L} (L_L)^C\,, \qquad \nu_{Ri} \to V_{ij} (\nu_{Rj})^C,  
\end{equation}
where $V$  is a $2\times  2$ dimensional  unitary matrix.  Notice that
the transformations given  by    Eq.\ (\ref{genCP}) satisfy  the    SM
symmetry  of  the mass-independent, conformal   invariant part  of the
Lagrangian;  only $M^\nu$  breaks this  symmetry  softly.  In such  an
approach,\cite{BRV,BBG}   one  looks    for all  possible   weak-basis
independent  combinations that can be  formed  by Yukawa couplings and
the  neutrino mass matrix   $M^\nu$, and are  simultaneously invariant
under the   transformations (\ref{genCP}). In   this way, we  find the
condition
\begin{equation}
\label{CPinv}
{\rm Im}\, \mbox{Tr} ( h^\dagger h M^{\nu\dagger} M^\nu M^{\nu\dagger}
h^T h^* M^\nu )\ =\ m_{N_1}m_{N_2} (m^2_{N_1} - m^2_{N_2})\,
{\rm Im} (h_{l1}h^*_{l2})^2\ =\ 0\, ,
\end{equation}
where $h  = (h_{l1}, h_{l2})$ is a  row vector that contains the Higgs
Yukawa couplings defined  in the mass basis  of  isosinglet neutrinos. 
{}From Eq.\ (\ref{CPinv}),  one  readily observes that CP   invariance
holds  if $m_{N_1}=m_{N_2}$ and/or  one of the isosinglet neutrinos is
massless.  These considerations may  be extended to models with  $n_L$
weak isodoublets and $n_R$ neutral  isosinglets.  In this case,  there
exist many conditions  analogous to Eq.\ (\ref{CPinv}),  which involve
high-order terms in the Yukawa-coupling  matrix $h$.  However, not all
of the conditions are sufficient and necessary for  CP invariance.  If
we assume that Higgs triplets are not present in the theory, the total
number of all non-trivial CP-violating phases  is ${\cal N}_{CP} = n_L
(n_R-1)$.\cite{KPS}

\begin{figure}

\begin{center}
\begin{picture}(360,300)(0,0)
\SetWidth{0.8}

\ArrowLine(0,270)(30,270)\ArrowLine(30,270)(60,270)\Line(60,270)(60,230)
\ArrowLine(90,230)(60,230)\DashArrowLine(60,270)(90,290){5}
\DashArrowLine(30,270)(60,230){5}\Text(60,250)[]{{\boldmath $\times$}}
\Text(0,265)[lt]{$N_i$}\Text(45,280)[]{$l$}\Text(65,250)[l]{$N_j$}
\Text(95,290)[l]{$\chi^-$}\Text(95,230)[l]{$l$}\Text(40,250)[r]{$\chi^+$}
\Text(50,210)[]{\bf (a)}

\ArrowLine(120,270)(150,270)\ArrowLine(150,270)(180,270)
\ArrowLine(180,270)(180,230)\ArrowLine(180,230)(210,230) 
\DashArrowLine(180,270)(210,290){5}\DashArrowLine(150,270)(180,230){5}
\Text(120,265)[lt]{$N_i$}\Text(165,280)[]{$\nu_l$}\Text(185,250)[l]{$N_j$}
\Text(215,290)[l]{$\chi^0$}\Text(215,230)[l]{$\nu_l$}
\Text(170,240)[r]{$\chi^0,H$}
\Text(170,210)[]{\bf (b)}

\ArrowLine(240,270)(270,270)\ArrowLine(270,270)(300,270)
\ArrowLine(300,270)(300,230)\ArrowLine(300,230)(330,230) 
\DashArrowLine(300,270)(330,290){5}\DashArrowLine(270,270)(300,230){5}
\Text(240,265)[lt]{$N_i$}\Text(285,280)[]{$\nu_l$}\Text(305,250)[l]{$N_j$}
\Text(335,290)[l]{$H$}\Text(335,230)[l]{$\nu_l$}
\Text(290,240)[r]{$\chi^0,H$}
\Text(290,210)[]{\bf (c)}

\DashArrowLine(0,150)(30,150){5}\ArrowArc(50,150)(20,0,180)
\ArrowArc(50,150)(20,180,360)\DashArrowLine(70,150)(100,150){5}
\Text(0,155)[bl]{$\chi^-$} \Text(100,155)[br]{$\chi^-$}
\Text(50,175)[b]{$N_i$}\Text(50,125)[t]{$l$}
\Text(50,100)[]{\bf (d)}

\DashArrowLine(120,150)(150,150){5}\ArrowArc(170,150)(20,0,180)
\ArrowArc(170,150)(20,180,360)\DashArrowLine(190,150)(220,150){5}
\Text(120,155)[bl]{$\chi^0$} \Text(220,155)[br]{$\chi^0$}
\Text(170,175)[b]{$N_i$}\Text(170,125)[t]{$\nu_l$}
\Text(170,100)[]{\bf (e)}

\DashArrowLine(240,150)(270,150){5}\ArrowArc(290,150)(20,0,180)
\ArrowArc(290,150)(20,180,360)\DashArrowLine(310,150)(340,150){5}
\Text(240,155)[bl]{$H$} \Text(340,155)[br]{$H$}
\Text(290,175)[b]{$N_i$}\Text(290,125)[t]{$\nu_l$}
\Text(290,100)[]{\bf (f)}

\ArrowLine(0,40)(30,40)\ArrowLine(30,40)(70,40)
\ArrowLine(70,40)(100,40)\DashArrowArc(50,40)(20,0,180){5}
\Text(0,45)[bl]{$l'$}\Text(100,45)[br]{$l$}
\Text(50,65)[b]{$\chi^+$}\Text(50,35)[t]{$N_i$}
\Text(50,0)[]{\bf (g)}

\ArrowLine(120,40)(150,40)\ArrowLine(150,40)(190,40)
\ArrowLine(190,40)(220,40)\DashArrowArc(170,40)(20,0,180){5}
\Text(120,45)[bl]{$\nu_{l'}$}\Text(220,45)[br]{$\nu_l$}
\Text(170,65)[b]{$\chi^0,H$}\Text(170,35)[t]{$N_i$}
\Text(170,0)[]{\bf (h)}

\ArrowLine(240,40)(270,40)\ArrowLine(270,40)(310,40)
\ArrowLine(310,40)(340,40)\DashArrowArc(290,40)(20,0,180){5}
\Text(240,45)[bl]{$N_j$}\Text(340,45)[br]{$N_i$}
\Text(290,65)[b]{$\chi^\mp,\chi^0,H$}\Text(290,35)[t]{$l^\mp,\nu_l,\nu_l$}
\Text(290,0)[]{\bf (j)}

\end{picture}\\[0.7cm]
\end{center}

\fcaption{One-loop graphs contributing to the renormalization of the 
couplings $\chi^-lN_i$, $\chi^0\nu_l N_i$ and $H\nu_lN_i$.}\label{fig:1}

\end{figure}                                  

\newpage

\setcounter{section}{4}
\setcounter{equation}{0}
\section{Renormalization}\label{sec:4}
\noindent 
At the  tree  level, CP violation   in particle decays amounts to  CPT
violation   and  therefore   vanishes   identically.   A non-vanishing
contribution  to CP  asymmetries  only arises  at  the one-loop level,
considering the diagrams shown in Fig.\ \ref{fig:1}.  For this reason,
it is important  to discuss  how  one-loop renormalization applies  to
heavy Majorana-neutrino models\cite{KP}  and its possible consequences
on CP asymmetries.

We start our discussion by expressing all bare quantities in terms of
renormalized ones in the following way:
\begin{eqnarray}
\label{Rbare}
\nu^0_{lL} &=& \sum\limits_{l'=1}^{n_L}\, \Big( \delta_{ll'}\, +\, 
\frac{1}{2}\delta Z^\nu_{ll'} \Big) \nu_{l'L}\, ,\qquad
l^0_L\ =\ \sum\limits_{l'=1}^{n_L}\, \Big( \delta_{ll'}\, +\, 
\frac{1}{2}\delta Z^l_{ll'} \Big) l'_L\, ,\\
N^0_i &=& \sum\limits_{j=1}^{n_R}\, \Big( \delta_{ij}\, +\, 
\frac{1}{2}\delta Z^N_{ij} \Big) N_j\, ,\quad
\tilde{\Phi} \ =\ \Big( 1\, +\, \frac{1}{2}\delta Z_\Phi \Big) \tilde{\Phi}
\, ,\quad
h^0_{lj} \ =\ h_{lj}\ +\ \delta h_{lj}\, ,\nonumber 
\end{eqnarray}
where unrenormalized kinematic parameters  and fields are indicated by
a superscript `0'.   Note that $\delta Z_\Phi$ collectively represents
the wave-function renormalization constants  of all components  of the
Higgs doublet $\tilde{\Phi}$ (or $\Phi$), i.e.\ the fields $\chi^\pm$,
$\chi^0$   and $H$.  In Appendix A,   we give analytic expressions for
Higgs and fermion self-energies. {}From these, one can easily see that
the   divergent  part of the  Higgs   wave-function renormalization is
exactly the same. In fact,  $\delta Z_\Phi$ is  universal in the limit
$M_H\to 0$.

Let us   now    consider  that   all  quantities in    the  Lagrangian
(\ref{LYint}) are bare and we  can substitute Eqs.\ (\ref{Rbare}) into
that bare  Lagrangian.   In addition to  the  renormalized Lagrangian,
which has the  same structural form as the  bare one, we then find the
counter-term (CT) Lagrangian
\begin{equation}
\label{deltaLY}
-\, \delta{\cal L}_Y\  =\ \frac{1}{2}\, \sum\limits_{l=1}^{n_L} h_{lj}
\sum\limits_{j=1}^{n_R}\, \Big(\, 2\frac{\delta h_{lj}}{h_{lj}}\, +\,
\delta Z_\Phi\, +\, \sum\limits_{l'=1}^{n_L} \delta Z^L_{l'l}\, +\,
 \sum\limits_{k=1}^{n_R} \delta Z^N_{jk}\, \Big)
\, \bar{L}_{l'}\tilde{\Phi} N_k\ +\ \mbox{H.c.},  
\end{equation}
where $L_l=(\nu_{lL},\ l_L )^T$ and $\delta  Z^L = (\delta Z^l, \delta
Z^\nu)$. Owing to charge and hypercharge conservation on the vertices,
it is not difficult to show by  naive power-counting that the one-loop
vertex corrections  in Fig.\ \ref{fig:1}(a)--(c) are ultra-violet (UV)
finite (see also Appendix A).

Despite the fact that vertex corrections are  UV finite by themselves,
the wave-function renormalizations of the Higgs and lepton isodoublets
and that of  neutrino isosinglets contain  UV divergences that do  not
cancel.  In accordance with the CT Lagrangian (\ref{deltaLY}), one may
require  that all UV  terms are to  be absorbed into the definition of
$h_{lj}$, i.e.
\begin{equation}
\label{deltah}
\delta h_{lj}\ =\ -\, \frac{1}{2}\, \Big(\, h_{lj}\delta Z^{\rm div}_{\Phi}\,
+\, \sum\limits_{l'=1}^{n_L} h_{l'j}\delta Z^{L*}_{l'l}\, +\,
\sum\limits_{k=1}^{n_R} h_{lk}\delta Z^N_{kj}\, \Big)\, .
\end{equation}
It  is important to  stress that one-loop renormalization involves the
dispersive    parts of  self-energies   and   effectively  leads to  a
redefinition  of  the   kinematic  parameters, whereas all  absorptive
corrections   remain unaffected.  Even   though   there might be  some
high-order dependence due  to the choice  of different renormalization
schemes, we carry  out the mass renormalization  in the  on-shell (OS)
scheme.\cite{OS} As we will see  in Section 7,   this scheme has  some
field-theoretic   advantages over other  schemes,  when applied to the
resummation approach describing the mixing of two unstable particles.

\setcounter{section}{5}
\setcounter{equation}{0}
\section{Resummation approach to unstable-particle mixing}\label{sec:5}
\noindent
The   consistent description    of   unstable   particles within   the
conventional framework of   perturbative S-matrix theory is an   issue
related to a  number  of field-theoretic difficulties. Since  unstable
particles  decay  exponentially  with  time,  they  cannot   appear as
asymptotic {\em in}  or {\em  out} states  in a process.  Furthermore,
finite-order  perturbation theory  breaks down.   The usual propagator
describing the unstable particle in the  intermediate state of a given
process displays a physical singularity when the particle comes on its
mass shell. One is therefore compelled to use resummation methods that
treat unstable particles and unstable-particle  mixing in a consistent
way; this  is  a  rather subtle issue   within the  context  of  gauge
theories.\cite{PP,ANPB}

In  a  simple scalar  theory    with one  unstable  particle,  Veltman
\cite{Velt} was able  to show that,  even if one removes  the unstable
particle from the initial and final states and substitutes it in terms
of asymptotic  states,  the  so-truncated S-matrix theory  will  still
maintain the  field-theoretic properties of  unitarity and  causality. 
Veltman's  truncated S-matrix  theory  is  rather  useful to  describe
resonant processes in collider experiments where the initial and final
states can be well   prepared and detected.  However,   this formalism
cannot directly be applied to the early Universe, as  it does not take
account  of the   many decoherentional  collisions   that an  unstable
particle may undergo  with  the thermal background  before  it decays. 
Therefore,   one must seek    a   method   that  isolates  the    {\em
  incoherent}\cite{LES}   part  of an   S-matrix  amplitude.  The  new
resummation method  should include finite width  effects in the mixing
and  decay of unstable  particles.  This will  be done in an effective
manner,   by   employing   a       procedure   related    to       the
Lehmann--Symanzik--Zimmermann  formalism (LSZ).\cite{LSZ} Then,    the
{\em  incoherent} decay   amplitude    derived with this method    may
equivalently be embedded into  a transition element  \cite{PP,ANPB} in
line with Veltman's  S-matrix formulation.  As we  will see in Section
9, the squared resummed decay amplitudes thus obtained will become the
relevant   collision terms  entering the  Boltzmann  equations for the
thermodynamic evolution of the Universe.

\begin{figure}

\begin{center}
\begin{picture}(360,100)(0,0)
\SetWidth{0.8}

\Text(0,60)[l]{$S_{i,\dots}$}\Text(30,60)[]{$=$}
\Text(60,60)[]{$\lim\limits_{p^2\to M^2_i}$}
\GOval(150,60)(30,20)(0){0.75}\Line(170,60)(200,60)\GCirc(215,60){15}{0.75}
\Line(230,60)(260,60)\Vertex(260,60){2}
\Line(120,90)(135,80)\Line(120,30)(135,40)\Vertex(120,90){2}
\Vertex(120,30){2}\Vertex(113,60){2}\Vertex(115,75){2}\Vertex(115,45){2}
\Text(115,90)[r]{$S_{i_1}$}\Text(115,30)[r]{$S_{i_n}$}
\Text(185,65)[b]{$S_k$}\Text(245,65)[b]{$S_j$}
\Text(270,80)[l]{$Z^{-1/2T}_{ji}\, \hat{\Delta}^{-1}_{ii}(p^2)$}
\LongArrow(250,50)(235,50)\Text(255,50)[l]{$p$}

\end{picture}\\[0.7cm]
\end{center}

\fcaption{Diagrammatic representation of the renormalized
  $n-1$-non-amputated amplitude, $S_{i,\dots}$, and the LSZ reduction
  formalism.}\label{fig:2}

\end{figure}

We  shall now    demonstrate  the effective  resummation   approach to
unstable particle mixing.  Let us consider  a theory with two  neutral
unstable scalars,  e.g.\ $S_1$ and  $S_2$.   The approach  can then be
extended to  the  case of   unstable fermions such  as  heavy Majorana
neutrinos.    The bare    (unrenormalized) fields  $S^0_i$   and their
respective    masses  $M^0_i$ may   then  be  expressed   in terms  of
renormalized fields $S_i$ and masses $M_i$ in the following way:
\begin{eqnarray}
\label{RenS0}
S^0_i & = & Z^{1/2}_{ij}\, S_j \ =\ \Big( \delta_{ij}\, +\,
\frac{1}{2} \delta Z_{ij}\Big) S_j\ ,\\
\label{RenMass}
(M^0_i)^2 & = & M^2_i\, +\, \delta M^2_i\ .
\end{eqnarray}
Here  and henceforth,  summation is  understood  over repeated indices
that  do not   appear  on  both  sides   of  an equation.   In   Eqs.\ 
(\ref{RenS0}) and (\ref{RenMass}), $Z^{1/2}_{ij}$ and $\delta M_i$ are
the wave-function  and  mass renormalization constants,  respectively,
which can be determined from renormalization conditions imposed on the
two-point  correlation functions, $\Pi_{ij}(p^2)$, for the transitions
$S_j\to S_i$  in some physical  scheme, such as the on-mass-shell (OS)
renormalization scheme.\cite{OS}  More  details may  be  found  in the
appendix.

In order to include the mixing of the unstable  scalars, we must first
calculate  all the $S_iS_j$  Green functions, with  $i,j =1,2$.  After
summing up  a geometric series  of the  self-energies $\Pi_{ij}(p^2)$,
the full propagators  may    be obtained by inverting   the  following
inverse propagator matrix:
\begin{equation}
\label{InvD12}
\Delta^{-1}_{ij} (p^2)\ =\ 
\left[
\begin{array}{cc}
p^2\, -\, (M^0_1)^2\, +\, \Pi_{11}(p^2) & \Pi_{12}(p^2)\\
\Pi_{21}(p^2) & p^2\, -\, (M^0_2)^2\, +\, \Pi_{22}(p^2)
\end{array} \right]\, .
\end{equation}
The result of inverting the matrix in Eq.\ (\ref{InvD12}) may be given by
\begin{eqnarray} 
\label{D11}
\Delta_{11}(p^2) &=& \left[ \, p^2\, -\, (M^0_1)^2
+\Pi_{11}(p^2)-\, \frac{\Pi^2_{12}(p^2)}{p^2-(M^0_2)^2+ 
\Pi_{22}(p^2)}\right]^{-1}\,
,\\
\label{D22}
\Delta_{22}(p^2) &=& \left[ \, p^2\, -\, (M^0_2)^2
+\Pi_{22}(p^2)-\, \frac{\Pi^2_{12}(p^2)}{p^2-(M^0_1)^2+ 
\Pi_{11}(p^2)}\right]^{-1}\,
,\\
\label{D12}
\Delta_{12}(p^2) &=& \Delta_{21}(p^2)\ =\
-\, \Pi_{12}(s) \Bigg[ \Big(p^2-(M^0_2)^2+\Pi_{22}(p^2)\Big)\nonumber\\
&&\times \Big( p^2 - (M^0_1)^2 +\Pi_{11}(p^2)\Big)\,
-\, \Pi^2_{12}(p^2)\, \Bigg]^{-1}\, .
\end{eqnarray}
where $\Pi_{12}(p^2)=\Pi_{21}(p^2)$.  Moreover, we observe the crucial
factorization  property  for   the  off-diagonal  ($i\not=j$) resummed
scalar propagators
\begin{eqnarray}
\label{Drel}
\Delta_{ij}(p^2) &=& -\, \Delta_{ii}(p^2)\
\frac{\Pi_{ij}(p^2)}{p^2\, -\, (M^0_j)^2\, +\, \Pi_{jj}(p^2)}\nonumber\\  
&=& -\, \frac{\Pi_{ij}(p^2)}{p^2\, -\, (M^0_i)^2\, +\,
\Pi_{ii}(p^2)}\ \Delta_{jj}(p^2)\ .
\end{eqnarray} 
The resummed unrenormalized scalar propagators $\Delta_{ij}(p^2)$ are
related to the respective renormalized ones $\hat{\Delta}_{ij}(p^2)$
through the expression
\begin{equation} 
\label{D_Dhat}
\Delta_{ij}(p^2)\ =\ 
Z^{1/2}_{im}\, \hat{\Delta}_{mn}(p^2)\, Z^{1/2T}_{nj}\, ,
\end{equation}
where  $\hat{\Delta}_{ij}(p^2)$     may  be    obtained    from  Eqs.\ 
(\ref{D11})--(\ref{D12}), just  by replacing  $M^0_i$  with $M_i$  and
$\Pi_{ij}(p^2)$    with  $\widehat{\Pi}_{ij}(p^2)$.    Note  that  the
property given in   Eq.\  (\ref{Drel}) will   also hold  true for  the
renormalized scalar propagators $\hat{\Delta}_{ij}(p^2)$.

Suppose that  we wish to find the   effective resummed decay amplitude
$\widehat{\cal  T}_{S_i}$ for the  decay   $S_i$ to  $n$ light  stable
scalars $S_{i_1}$, \dots, $S_{i_n}$.  In analogy to the LSZ formalism,
one starts with the Green function  describing the transition shown in
Fig.\  \ref{fig:2}, and amputates  the  external legs by their inverse
propagators.  For  the    stable  external  lines   $S_{i_1}$,  \dots,
$S_{i_n}$, the procedure is  essentially the same  with the usual  LSZ
formalism.  This formalism may then  be extended to the external  line
describing the  $S_i   S_j$ system.  The  intermediate   steps of this
procedure are given by
\begin{eqnarray}
\label{LSZ2}
\widehat{\cal T}_{i\dots}& = & \lim\limits_{p^2\to M^2_i}\ 
{\cal T}^{amp}_{k\dots}  
Z^{1/2}_{km}\, \hat{\Delta}_{mn}(p^2)\, Z^{1/2T}_{nj}
Z^{-1/2T}_{ji} \hat{\Delta}^{-1}_{ii}(p^2) \nonumber\\
&=& \lim\limits_{p^2\to M^2_i}\Big[ {\cal T}^{amp}_{k\dots}Z^{1/2}_{ki}\ 
-\ {\cal T}^{amp}_{k\dots}Z^{1/2}_{km} \frac{\widehat{\Pi}_{mi}(p^2) 
(1-\delta_{mi})}{p^2-M^2_m+\widehat{\Pi}_{mm}(p^2)}\, \Big]\nonumber\\
&=& {\cal T}_{i\dots}\ -\ {\cal T}_{j\dots}\frac{\widehat{\Pi}_{ji}(M^2_i) 
(1-\delta_{ij})}{M^2_i-M^2_j+\widehat{\Pi}_{jj}(M^2_i)}\  ,
\end{eqnarray}
where ${\cal T}_{i\dots}$ and ${\cal T}_{j\dots}$ are the renormalized
transition elements   evaluated in the  stable-particle approximation. 
One  should bear  in   mind that   the  OS  renormalized self-energies
$\widehat{\Pi}_{ji}(M^2_i)$   in Eq.\ (\ref{LSZ2})  have  no vanishing
absorptive  parts, as renormalization can  only  modify the dispersive
(real) part   of these  self-energies.    The reason  is  that  the CT
Lagrangian must be Hermitian as opposed to  the absorptive parts which
are anti-Hermitian. In fact, these additional width mixing effects are
the ones we wish to include in our formalism  for decay amplitudes and
are absent in  the   conventional perturbation  theory.  It   is  also
important  to  observe that our approach   to  decays is not singular,
i.e.\ $\widehat{S}_{i\dots}$  displays  an  analytic behaviour  in the
degenerate limit  $M^2_i\to M^2_j$, because  of the appearance  of the
imaginary term $i{\rm Im}\widehat{\Pi}_{jj}(M^2_i)$ in the denominator
of the   mixing  factor   present  in  the   last  equality   of  Eq.\ 
(\ref{LSZ2}).  Finally,  we  must stress that  the  inclusion of these
phenomena  has been  performed  in  an  effective  manner.  Since  the
decaying  unstable  particle       cannot appear   in    the   initial
state,\cite{Velt}  the resummed  decay  amplitude must  be regarded as
being a  part which can    effectively  be embedded  into a   resummed
S-matrix  element.\cite{PP} This  resummed S-matrix element  describes
the dynamics of the very same unstable particle,  which is produced by
some asymptotic  states,   resides  in  the  intermediate  state,  and
subsequently decays  either  directly or   indirectly, through mixing,
into the observed final states.

The resummation  approach outlined   above can now  carry  over  to the
mixing between  two unstable fermions, call them  $f_1$ and $f_2$.  As
we did  for  the  case of  scalars,   we express  the  bare  left- and
right-handed chiral  fields, $f^0_{Li}$ and $f^0_{Ri}$ (with $i=1,2$),
in terms of renormalized fields as follows:
\begin{equation}
f^0_{Li}\ =\ Z^{1/2}_{Lij}\, f_{Lj}\ , \quad\qquad
f^0_{Ri}\ =\ Z^{1/2}_{Rij}\, f_{Rj}\ ,
\end{equation}
where  $Z^{1/2}_{Lij}$     ($Z^{1/2}_{Rij}$) is     the  wave-function
renormalization constant for the left  (right)- handed chiral  fields,
which may  be  determined from the   fermionic self-energy transitions
$f_j\to   f_i$,  $\Sigma_{ij}  (\not\!\!  p)$,      e.g.\ in the    OS
renormalization scheme.\cite{KP}   Analogously to Eq.\ (\ref{InvD12}),
the resummed fermion propagator matrix may be obtained from
\begin{equation}
\label{InvS}
S_{ij}(\not\! p)\ =\ \left[ \begin{array}{cc} \not\! p - m^0_1 +
\Sigma_{11}(\not\! p) & \Sigma_{12}(\not\! p)\\
\Sigma_{21}(\not\! p) & \not\! p - m^0_2 + \Sigma_{22}(\not\! p)
\end{array} \right]^{-1},\qquad
\end{equation}
where $m^0_{1,2}$ are the bare fermion masses, which can be decomposed
into the   OS  renormalized masses  $m_{1,2}$ and  the   CT mass terms
$\delta m_{1,2}$  as  $m^0_{1,2}=m_{1,2} +  \delta m_{1,2}$. Inverting
the matrix-valued $2\times 2$ matrix in Eq.\ (\ref{InvS}) yields
\begin{eqnarray}
\label{S11}
S_{11}(\not\! p) &=& \Big[\not\! p\, -\, m^0_1\, +\,
\Sigma_{11}(\not\! p)\, -\, \Sigma_{12}(\not\! p)
\frac{1}{\not\! p - m^0_2 + \Sigma_{22}(\not\! p)}
\Sigma_{21}(\not\! p) \Big]^{-1}, \\
\label{S22}
S_{22}(\not\! p) &=& \Big[\not\! p\, -\, m^0_2\, +\,
\Sigma_{22}(\not\! p)\, -\, \Sigma_{21}(\not\!
p) \frac{1}{\not\! p - m^0_1 + \Sigma_{11}(\not\! p)}
\Sigma_{12}(\not\! p) \Big]^{-1}, \\
\label{S12}
S_{12}(\not\! p) &=& -\, S_{11}(\not\! p)\,
\Sigma_{12}(\not\! p)\, \Big[ \not\! p\, -\, m^0_2\, +\,
\Sigma_{22}(\not\! p) \Big]^{-1} \nonumber\\ &=& -\, \Big[
\not\! p\, -\, m^0_1\, +\, \Sigma_{11}(\not\! p) \Big]^{-1}
\Sigma_{12}(\not\! p)\, S_{22}(\not\! p)\ ,\\
\label{S21}
S_{21}(\not\! p) &=& -\, S_{22}(\not\! p)\,
\Sigma_{21}(\not\! p)\, \Big[ \not\! p\, -\, m^0_1\, +\,
\Sigma_{11}(\not\! p) \Big]^{-1}\, \nonumber\\ &=& -\, \Big[
\not\! p\, -\, m^0_2\, +\, \Sigma_{22}(\not\! p)
\Big]^{-1} \Sigma_{21}(\not\! p)\, S_{11}(\not\! p)\ .
\end{eqnarray}
Equations  (\ref{S12})  and  (\ref{S21})   show   that  the   resummed
propagators   $S_{12}(\not\!\!   p)$  and   $S_{21}(\not\!\!   p)$ are
endowed with a factorization property  analogous to Eq.\ (\ref{Drel}). 
Similarly,  the renormalized  and  unrenormalized resummed propagators
are related by
\begin{equation}
\label{S_Shat}
S_{ij}(\not\! p)\ =\ (Z^{1/2}_{Lim}\, P_L\ +\ Z^{1/2}_{Rim}\, P_R )\,
\widehat{S}_{mn}(\not\! p)\, (Z^{1/2\dagger}_{Lnj}\, P_R\ +\ 
Z^{1/2\dagger}_{Rnj}\, P_L )\, ,
\end{equation}
where the caret on $S_{ij}(\not\!\!   p)$ indicates that the  resummed
fermionic  propagators have   been   renormalized in  the OS  scheme.  
Moreover, the renormalized  propagators $\widehat{S}_{ij}$ $(\not\!\!  
p)$    may   be   obtained    by   $S_{ij}(\not\!\!   p)$    in  Eqs.\ 
(\ref{S11})--(\ref{S21}), if the  obvious replacements  $m^0_i\to m_i$
and  $\Sigma_{ij}(\not\!  p)\to  \widehat{\Sigma}_{ij}(\not\!  p)$ are
made.

By  analogy, one   can     derive  the  resummed    decay   amplitude,
$\widehat{\cal T}_{i\dots}$,   of  the unstable fermion    $f_i\to X$,
as we did for the scalar case.  More explicitly, we have
\begin{eqnarray}
\label{LSZ3}
\widehat{\cal T}_{i\dots}\, u_i(p) &=& 
{ \cal T}^{amp}_{k\dots}\, (Z^{1/2}_{Lkm}\, P_L\, +\, Z^{1/2}_{Rkm}\, P_R )\,
\widehat{S}_{mn}(\not\! p)\, (Z^{1/2\dagger}_{Lnj}\, P_R\, +\, 
Z^{1/2\dagger}_{Rnj}\, P_L )\nonumber\\
&&\times (Z^{-1/2\dagger}_{Lji}\, P_R\, +\, 
Z^{-1/2\dagger}_{Rji}\, P_L )\, \widehat{S}^{-1}_{ii}(\not\! p)\, 
u_i(p)\\
&=& {\cal T}_{i\dots}\,  u_i(p)\ -\ (1-\delta_{ij}) {\cal T}_{j\dots}\, 
\widehat{\Sigma}_{ji}(\not\! p)\, \Big[ \not\! p\, -\, m_j\, +\,
\widehat{\Sigma}_{jj}(\not\! p) \Big]^{-1} u_i(p)\, .\nonumber
\end{eqnarray}
Again,   ${\cal  T}_{i\dots}$  represent  the respective  renormalized
transition amplitudes evaluated  in the stable-particle approximation. 
The  amplitudes   ${\cal  T}_{i\dots}$   also include all   high-order
$n$-point functions, such as  vertex corrections. Based on the formula
(\ref{LSZ3}),  we shall calculate the CP  asymmetries in the decays of
heavy Majorana neutrinos in the next section.

\setcounter{section}{6}
\setcounter{equation}{0}
\section{CP asymmetries}\label{sec:6}
\noindent
The resummation  approach  presented in  the  previous  section may be
applied    to  describe  $\varepsilon$-  and  $\varepsilon'$-type   CP
violation   in  heavy   Majorana neutrino   decays    shown  in  Fig.\ 
\ref{fig:3}.  The same formalism  may also  be  used to  determine the
collision terms for the inverse decays, which occur in the formulation
of the Boltzmann equations (see also Section 8).

\begin{figure}

\begin{center}
\begin{picture}(300,100)(0,0)
\SetWidth{0.8}

\Vertex(50,50){2}
\Line(50,50)(90,50)\Text(70,62)[]{$N_i$}
\Line(130,50)(170,50)\Text(150,62)[]{$N_j$}
\GCirc(110,50){20}{0.9}\Text(110,50)[]{{\boldmath $\varepsilon$}}
\GCirc(180,50){10}{0.9}\Text(180,50)[]{\boldmath $\varepsilon$\bf'}
\DashArrowLine(187,55)(220,80){5}\Text(225,80)[l]{$\Phi^\dagger$}
\ArrowLine(187,45)(220,20)\Text(225,20)[l]{$L$}

\end{picture}\\
\end{center}

\fcaption{$\varepsilon$- and $\varepsilon'$-type CP violation in the
  decays of heavy Majorana neutrinos.}\label{fig:3}

\end{figure}

Let  us  consider the decay  $N_1\to  l^-\chi^+$ in   a model with two
right-handed neutrinos.  The inclusion of all  other decay channels is
then obvious.  We shall   first write  down the  transition  amplitude
responsible  for $\varepsilon$-type CP   violation, denoted as  ${\cal
  T}^{(\varepsilon)}_N$, and then take CP-violating vertex corrections
into account. Applying (\ref{LSZ3}) to heavy Majorana neutrino decays,
we obtain
\begin{equation}
\label{TN1eps}
{\cal T}^{(\varepsilon)}_{N_1}\ =\ h_{l1}\, \bar{u}_lP_R u_{N_1}\ -\
ih_{l2}\, \bar{u}_l P_R \Big[\not\! p - m_{N_2} + i\Sigma_{22}^{abs}
(\not\! p)\Big]^{-1} \Sigma_{21}^{abs}(\not\! p) u_{N_1}\, ,
\end{equation}
where the absorptive  part of the  one-loop transitions  $N_j\to N_i$,
with $i,j=1,2$, has the general form
\begin{equation}
\label{Sigabs}
\Sigma^{abs}_{ij} (\not\! p)\ =\ A_{ij}(p^2)\not\! p P_L\, +\, 
A^*_{ij}(p^2)\not\! p P_R\, ,
\end{equation}
with 
\begin{equation}
\label{Aij}
A_{ij}(p^2)\ =\ \frac{h_{l'i}h^*_{l'j}}{32\pi}\, \Big[\, \frac{3}{2}\, +\,
\frac{1}{2}\, \Big( 1-\frac{M^2_H}{p^2} \Big)^2\, \Big]\, .
\end{equation}
In the  limit $M_H\to 0$,  Eq.\  (\ref{Aij}) gives $A_{ij}  =  h_{l'i}
h^*_{l'j} / (16\pi)$.   The CP-transform resummed amplitude describing
the     decay  $N_1\to        l^+\chi^-$,             $\overline{{\cal
    T}}^{(\varepsilon)}_{N_1}$, reads
\begin{eqnarray}
\label{TCPN1eps}
\overline{{\cal T}}^{(\varepsilon)}_{N_1} &=& 
h^*_{l1}\, \bar{v}_{N_1}P_L v_l\ -\
ih_{l2}\, \bar{v}_{N_1} \Sigma_{12}^{abs}(-\not\! p) \Big[\, -\not\! p - 
m_{N_2} + i\Sigma_{22}^{abs} (-\not\! p)\Big]^{-1} P_L v_l\nonumber\\
&=& h^*_{l1}\, \bar{u}_lP_L u_{N_1}\ -\
ih^*_{l2}\, \bar{u}_l P_L  \Big[\not\! p - m_{N_2} + 
i\overline{\Sigma}_{22}^{abs}
(\not\! p)\Big]^{-1} \overline{\Sigma}_{21}^{abs}(\not\! p) u_{N_1}\, , 
\end{eqnarray}
where
\begin{equation}
\label{SigCabs}
\overline{\Sigma}^{abs}_{ij} (\not\! p)\ =\ A_{ij}(p^2)\not\! p P_R\, +\, 
A^*_{ij}(p^2)\not\! p P_L
\end{equation}
is the charge-conjugate absorptive self-energy.  The last step of Eq.\ 
(\ref{TCPN1eps}) is derived by making use of the identities
\begin{equation}
  \label{idCP}
u(p,s)\ =\ C\bar{v}^T(p,s)\, ,\qquad C\gamma_\mu C^{-1}\ =\ -\gamma^T_\mu\, .  
\end{equation}
The expressions in  Eqs.\  (\ref{TN1eps}) and (\ref{TCPN1eps}) may  be
simplified even further,  if the Dirac equation  of motion is employed
for the  external spinors. Then,  the  two resummed  decay amplitudes,
${\cal        T}^{(\varepsilon)}_{N_1}$     and       $\overline{{\cal
    T}}^{(\varepsilon)}_{N_1}$, take the simple form
\begin{eqnarray}
\label{TN}
{\cal T}^{(\varepsilon)}_{N_1} &=& \bar{u}_lP_R u_{N_1}\, 
\Big[\, h_{l1}\, -\, ih_{l2}\, \frac{m^2_{N_1}(1+iA_{22})A^*_{21}
+m_{N_1}m_{N_2}A_{21}}{m^2_{N_1}(1+iA_{22})^2 -m^2_{N_2}}\, \Big]\, ,\\
\label{TCPN}
\overline{{\cal T}}^{(\varepsilon)}_{N_1} &=&\bar{u}_lP_L u_{N_1}\, 
\Big[\, h^*_{l1}\, -\, ih^*_{l2}\, \frac{m^2_{N_1}(1+iA_{22})A_{21}
+m_{N_1}m_{N_2}A^*_{21}}{m^2_{N_1}(1+iA_{22})^2 -m^2_{N_2}}\, \Big]\, .
\end{eqnarray}
In addition, the respective transition amplitudes involving the decays
$N_2\to   l^-\chi^+$,  ${\cal  T}^{(\varepsilon)}_{N_2}$,  and $N_2\to
l^+\chi^-$,    $\overline{{\cal  T}}^{(\varepsilon)}_{N_2}$,  may   be
obtained by interchanging the indices `1' and  `2' everywhere in Eqs.\ 
(\ref{TN}) and (\ref{TCPN}).

In order    to  study   the   $\varepsilon$-   and $\varepsilon'$-type
mechanisms  of CP violation in    heavy Majorana neutrino decays,   we
define the following CP-violating quantities:
\begin{eqnarray}
\label{epsNi}
\varepsilon_{N_i} & =& \frac{|{\cal T}^{(\varepsilon)}_{N_i}|^2\, -\,
|\overline{{\cal T}}^{(\varepsilon)}_{N_i}|^2}{
|{\cal T}^{(\varepsilon)}_{N_i}|^2\, +\, 
|\overline{{\cal T}}^{(\varepsilon)}_{N_i}|^2}\ ,\qquad \mbox{for}\ i=1,2\, ,\\
\label{epsN}
\varepsilon_N & =& \frac{|{\cal T}^{(\varepsilon)}_{N_1}|^2\, +\,
|{\cal T}^{(\varepsilon)}_{N_2}|^2\,
-\, |\overline{{\cal T}}^{(\varepsilon)}_{N_1}|^2 
\, -\, |\overline{{\cal T}}^{(\varepsilon)}_{N_2}|^2}{
|{\cal T}^{(\varepsilon)}_{N_1}|^2\, +\, |{\cal T}^{(\varepsilon)}_{N_2}|^2
\, +\, |\overline{{\cal T}}^{(\varepsilon)}_{N_1}|^2\, +\,
|\overline{{\cal T}}^{(\varepsilon)}_{N_2}|^2}\ .
\end{eqnarray}
Correspondingly, the  CP-violating parameters $\varepsilon'_{N_i}$ and
$\varepsilon'_{N}$ may be defined by
\begin{eqnarray}
\label{epsNipr}
\varepsilon'_{N_i} & =& \frac{|{\cal T}^{(\varepsilon')}_{N_i}|^2\, -\,
|\overline{{\cal T}}^{(\varepsilon')}_{N_i}|^2}{
|{\cal T}^{(\varepsilon')}_{N_i}|^2\, +\, 
|\overline{{\cal T}}^{(\varepsilon')}_{N_i}|^2}\ ,
\qquad \mbox{for}\ i=1,2\, ,\\
\label{epsNpr}
\varepsilon'_N & =& \frac{|{\cal T}^{(\varepsilon')}_{N_1}|^2\, +\,
|{\cal T}^{(\varepsilon')}_{N_2}|^2\,
-\, |\overline{{\cal T}}^{(\varepsilon')}_{N_1}|^2 
\, -\, |\overline{{\cal T}}^{(\varepsilon')}_{N_2}|^2}{
|{\cal T}^{(\varepsilon')}_{N_1}|^2\, +\, |{\cal T}^{(\varepsilon')}_{N_2}|^2
\, +\, |\overline{{\cal T}}^{(\varepsilon')}_{N_1}|^2\, +\,
|\overline{{\cal T}}^{(\varepsilon')}_{N_2}|^2}\ .
\end{eqnarray}
The last parameters quantify  CP violation coming exclusively from the
one-loop   irreducible   vertices.     In  Eqs.\  (\ref{epsNi})    and
(\ref{epsN}), the  parameters $\varepsilon_{N_i}$  and $\varepsilon_N$
share the common property that they  do not depend  on the final state
that $N_i$ decays, despite the fact that the individual squared matrix
elements do.  In  general, both $\varepsilon$- and $\varepsilon'$-type
contributions are not  directly  distinguishable in the  decay  widths
$\Gamma    (N_i\to      l^\pm\chi^\pm)$, unless  $\varepsilon_{N_i}\gg
\varepsilon'_{N_i}$ and vice  versa, for some  range  of the kinematic
parameters.  Evidently, the physical CP asymmetries are given by
\begin{eqnarray}
\label{deltaNi}
\delta_{N_i} &=& \frac{\Gamma (N_i\to L\Phi^\dagger )\, -\, 
\Gamma (N_i\to L^C \Phi)}{\Gamma (N_i\to L\Phi^\dagger )\, +\, 
\Gamma (N_i\to L^C \Phi)}\ , \qquad \mbox{for}\ i=1,2\, ,\\
\label{deltaN}
\delta_N &=& \frac{\sum_{i=1}^2\Gamma (N_i\to L\Phi^\dagger )\, -\, 
\sum_{i=1}^2\Gamma (N_i\to L^C \Phi)}{\sum_{i=1}^2
\Gamma (N_i\to L\Phi^\dagger )\, +\, \sum_{i=1}^2\Gamma (N_i\to L^C \Phi)}\ , 
\end{eqnarray}
where $L$ refers to  all fermionic degrees  of freedom of the leptonic
isodoublet that heavy Majorana neutrinos  can decay. Nevertheless, the
parameters $\varepsilon_{N_i}$,  $\varepsilon_N$, $\varepsilon'_{N_i}$
and  $\varepsilon'_N$ defined above are  very  useful to determine the
contributions due to the different mechanisms of CP violation.

We now turn to the calculation of the CP-violating contribution, which
is      entirely due  to    the  heavy-neutrino   self-energy effects. 
Substituting Eqs.\ (\ref{TN}) and  (\ref{TCPN}) into (\ref{epsNi}), we
arrive at the simple formulas\cite{APRD}
\begin{eqnarray}
\label{epsN1}
\varepsilon_{N_1} &\approx& \frac{{\rm Im} ( h^*_{l1}h_{l2})^2}{
|h_{l1}|^2|h_{l2}|^2}\ 
\frac{\Delta m^2_N m_{N_1} \Gamma_{N_2} }{(\Delta m^2_N)^2\, +\, 
m^2_{N_1}\Gamma^2_{N_2}}\, ,\\
\label{epsN2}
\varepsilon_{N_2} &\approx& \frac{{\rm Im} ( h^*_{l1}h_{l2})^2}{
|h_{l1}|^2|h_{l2}|^2}\ 
\frac{\Delta m^2_N m_{N_2} \Gamma_{N_1} }{(\Delta m^2_N)^2\, +\, 
m^2_{N_2}\Gamma^2_{N_1}}\, ,
\end{eqnarray}
where  $\Delta m^2_N  = m^2_{N_1} -   m^2_{N_2}$ and 
\begin{equation}
  \label{GammaN}
\Gamma_{N_i}\ =\  \frac{|h_{li}|^2}{8\pi}\ m_{N_i}   
\end{equation}
are  the  decay  widths of  the  heavy Majorana  neutrinos.  Equations
(\ref{epsN1}) and (\ref{epsN2}) are a  very good approximation for any
range of   heavy-neutrino masses  of  interest.  Both   CP asymmetries
$\varepsilon_{N_1}$ and $\varepsilon_{N_2}$  are of the  same sign and
go individually to zero  when $\Delta m^2_N\to 0$, as  it should be on
account of   Eq.\ (\ref{CPinv}).   In the    conventional perturbation
theory, the width   terms $m^2_{N_1}  \Gamma^2_{N_2}$  and  $m^2_{N_2}
\Gamma^2_{N_1}$ occurring in the last denominators on the RHS of Eqs.\ 
(\ref{epsN1}) and (\ref{epsN2}) are   absent.  This very last fact  is
precisely what causes a singular behaviour when the degeneracy between
the   two heavy Majorana   neutrinos  is exact.   On physical grounds,
however,  the only   natural    parameter that can regulate    such  a
singularity  is the   finite width of  the heavy   neutrinos, which is
naturally implemented within the resummation approach.

{}From  Eqs.\ (\ref{epsN1}) and (\ref{epsN2}), it  is not difficult to
derive   the   sufficient  and   necessary   conditions   for resonant
enhancement of CP  violation. To be specific,  CP violation can  be of
order unity if and only if
\begin{eqnarray}
    \label{CPcond}
  {\rm  (i)}&&\hspace{-0.3cm} 
m_{N_1}\, -\, m_{N_2}\ \sim\ \pm\, A_{22} m_{N_2}\, 
         =\, \frac{\Gamma_{N_2}}{2}\
        \ \mbox{and/or}\quad A_{11} m_{N_1}\, =\,
         \frac{\Gamma_{N_1}}{2}\, ,\\
    \label{dCP}
  {\rm (ii)}&&\hspace{-0.3cm} 
 \delta_{CP}\ =\ \frac{|{\rm Im} (h^*_{l1}h_{l2})^2|}{|h_{l1}|^2
                         |h_{l2}|^2}\ \approx 1\ .
\end{eqnarray}

Before we present numerical  estimates of CP asymmetries, we calculate
for  completeness the contributions  to  CP violation arising entirely
from vertex effects.    The $\varepsilon'$-type contributions  can  be
significant for large differences  of heavy neutrino masses, e.g.\ for
$m_{N_1}-m_{N_2}\sim m_{N_1}$   or $m_{N_2}$.    In this  regime, both
$\varepsilon$-type  and $\varepsilon'$-type effects  are of comparable
order.\cite{LS1} It is useful to define first the function
\begin{equation}
\label{Fxa}
F(x,\alpha)\ =\ \sqrt{x}\, \Big[\, 1\, -\, \alpha\, -\, (1+x)
\ln\Big(\frac{1-\alpha+x}{x}\Big)\, \Big]\, .
\end{equation}
With $\alpha = 0$, $F(x,\alpha)$ reduces to the Fukugita-Yanagida loop
function $f(x) = \sqrt{x} [ 1 - (1+x) \ln  (1 + 1/x)]$.\cite{FY} Then,
$L$-violating absorptive  parts of the one-loop vertices $\chi^+lN_i$,
$\chi^0 \nu_l  N_i$    and    $H \nu_l    N_i$,   shown  in     Figs.\ 
\ref{fig:1}(a)--(c), are given by
\begin{eqnarray}
  \label{eps'lN}
{\cal V}^{abs}_{\chi^+lN_i}(\not\! p) &=& -\, \frac{h^*_{l'i}h_{l'j}h_{lj}}{
16\pi\sqrt{p^2}}\ \not\! p P_L\, F\Big(\frac{m^2_{Nj}}{p^2}\ ,0\Big)\, ,\\
  \label{eps'nuN}
{\cal V}^{abs}_{\chi^0\nu_lN_i}(\not\! p) &=& {\cal V}^{abs}_{H\nu_lN_i}
(\not\! p)\nonumber\\
&=& -\, \frac{h^*_{l'i}h_{l'j}h_{lj}}{
32\pi\sqrt{p^2}}\ \not\! p P_L\, \Big[\, F\Big(\frac{m^2_{Nj}}{p^2}\ ,0\Big)
\ +\  F\Big(\frac{m^2_{Nj}}{p^2}\ , \frac{M^2_H}{p^2}\Big)\, \Big]\, .\quad
\end{eqnarray}
Here, we   have assumed that    the external decaying   heavy Majorana
neutrinos are off-shell, whereas  the leptons and  Higgs fields are on
their mass shell. The complete  analytic expressions are calculated in
the  appendix.  Using  Eqs.\  (\ref{eps'lN}) and  (\ref{eps'nuN})  and
neglecting      wave-function    contributions,   we   compute     the
$\varepsilon'$-type  CP asymmetry  in  the  conventional  perturbation
theory.  Considering   all   decay channels   for  the  decaying heavy
Majorana neutrino, e.g.\ $N_1$, we find
\begin{eqnarray}
\label{eps'N1}
\varepsilon'_{N_1} &=& \frac{{\rm Im}(h^*_{l1}h_{l2})^2}{
16\pi |h_{l1}|^2\, [\frac{3}{4} +\frac{1}{4}(1-M^2_H/m^2_{N_1})^2]}\
\Big\{\, \frac{5}{4}\, F\Big(\frac{m^2_{N_2}}{m^2_{N_1}}\ ,0\Big)\, +\, 
\frac{1}{4}\, F\Big(\frac{m^2_{N_2}}{m^2_{N_1}}\ , 
\frac{M^2_H}{m^2_{N_1}}\Big)\nonumber\\
&&+\, \frac{1}{4}\, \Big( 1\, -\, \frac{M^2_H}{m^2_{N_1}}\Big)^2\,
\Big[\, F\Big(\frac{m^2_{N_2}}{m^2_{N_1}}\ ,0\Big)\, +\, 
F\Big(\frac{m^2_{N_2}}{m^2_{N_1}}\ , \frac{M^2_H}{m^2_{N_1}}\Big)
\, \Big]\, \Big\}\, .
\end{eqnarray}
In the  limit  $M_H\to 0$, the  last  formula simplifies  to the known
result\cite{FY,MAL,CEV,epsilonprime}
\begin{equation}
\label{eps'}
\varepsilon'_{N_1}\ =\ \frac{{\rm Im}(h^*_{l1}h_{l2})^2}{
8\pi |h_{l1}|^2 }\ f\Big(\frac{m^2_{N_2}}{m^2_{N_1}}\Big)\, .
\end{equation}
Unlike $\varepsilon_{N_1}$,   $\varepsilon'_{N_1}$ does not  vanish in
the degenerate limit  of the  two  heavy Majorana neutrinos $N_1$  and
$N_2$.    However, when the  value  of   $m_{N_1}$ approaches that  of
$m_{N_2}$, the  $\varepsilon'$-type  part of the  transition amplitude
squared for the $N_1$ decay becomes equal  but opposite in sign to the
respective  one  of    the $N_2$  decay.    As   a result,  these  two
$\varepsilon'$-type terms cancel one another, leading to the vanishing
of the  CP-violating   parameter   $\varepsilon'_N$ defined  in   Eq.\ 
(\ref{epsNpr}).  Consequently, as  opposed  to  $\varepsilon$ effects,
$\varepsilon'$ effects cannot become resonant for any kinematic region
of mass parameters.

Both  $\varepsilon$  and $\varepsilon'$  contributions can be included
into the  resummed decay amplitudes.  Considering Eqs.\ (\ref{eps'lN})
and (\ref{eps'nuN}) into account, we obtain
\begin{eqnarray}
\label{TN1}
{\cal T}_{N_1} \hspace{-2pt}&=&\hspace{-2pt} \bar{u}_l P_R\, \Big\{ 
h_{l1}+i{\cal V}^{abs}_{l1} (\not\! p) -
i\Big[h_{l2}+i{\cal V}^{abs}_{l2} (\not\! p) \Big]\nonumber\\
&&\times \Big[\not\! p - m_{N_2} + i\Sigma_{22}^{abs}(\not\! p)\Big]^{-1} 
\Sigma_{21}^{abs}(\not\! p)\Big\} u_{N_1} ,\\
\label{TCPN1}
\overline{{\cal T}}_{N_1} \hspace{-2pt}&=&\hspace{-2pt} 
\bar{u}_l P_L\, \Big\{ 
h^*_{l1}+i\overline{{\cal V}}^{abs}_{l1} (\not\! p) 
 - i\Big[h^*_{l2}+i\overline{{\cal V}}^{abs}_{l2} (\not\! p) \Big]\nonumber\\ 
&&\times 
\Big[\not\! p - m_{N_2} + i\overline{\Sigma}_{22}^{abs}(\not\! p)\Big]^{-1} 
\overline{\Sigma}_{21}^{abs}(\not\! p)\Big\} u_{N_1} ,
\end{eqnarray}
where  the notation of  the    off-shell one-loop vertices has    been
simplified to ${\cal V}^{abs}_{li}(\not\!\! p)$.  The vertex functions
$\overline{{\cal V}}^{abs}_{li}(\not\!  p)$ are the  charge conjugates
of  ${\cal  V}^{abs}_{li}(\not\! p)$  and may  hence be recovered from
Eqs.\ (\ref{eps'lN})  and   (\ref{eps'nuN}),  by taking  the   complex
conjugate  for  the Yukawa couplings  and replacing  $P_R$ with $P_L$. 
Although   the calculation     of   the   CP-violating     observables
$\delta_{N_i}$   defined    in    Eq.\  (\ref{deltaNi})     is   quite
straightforward  from Eqs.\ (\ref{TN1})  and  (\ref{TCPN1}), it is not
very easy to present analytic expressions in a compact form. 

\begin{figure}[hb]
   \leavevmode
 \begin{center}
   \epsfxsize=11.cm
   \epsffile[0 0 539 652]{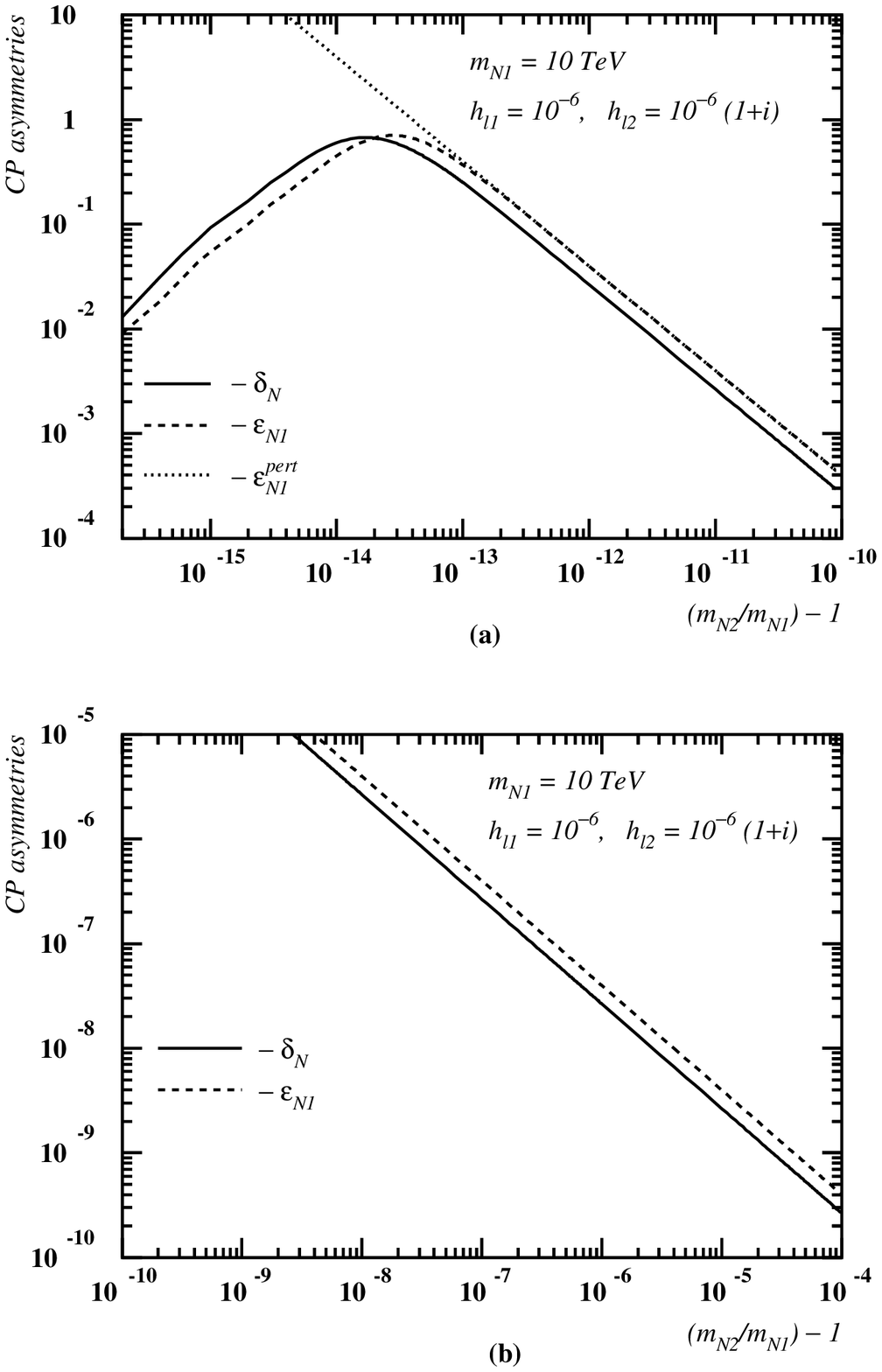}
 \end{center}
 
\fcaption{Numerical estimates of CP asymmetries in scenario
   I.}\label{fig:4}

\end{figure}

\begin{figure}[ht]
   \leavevmode
 \begin{center}
   \epsfxsize=11.cm
   \epsffile[0 0 539 652]{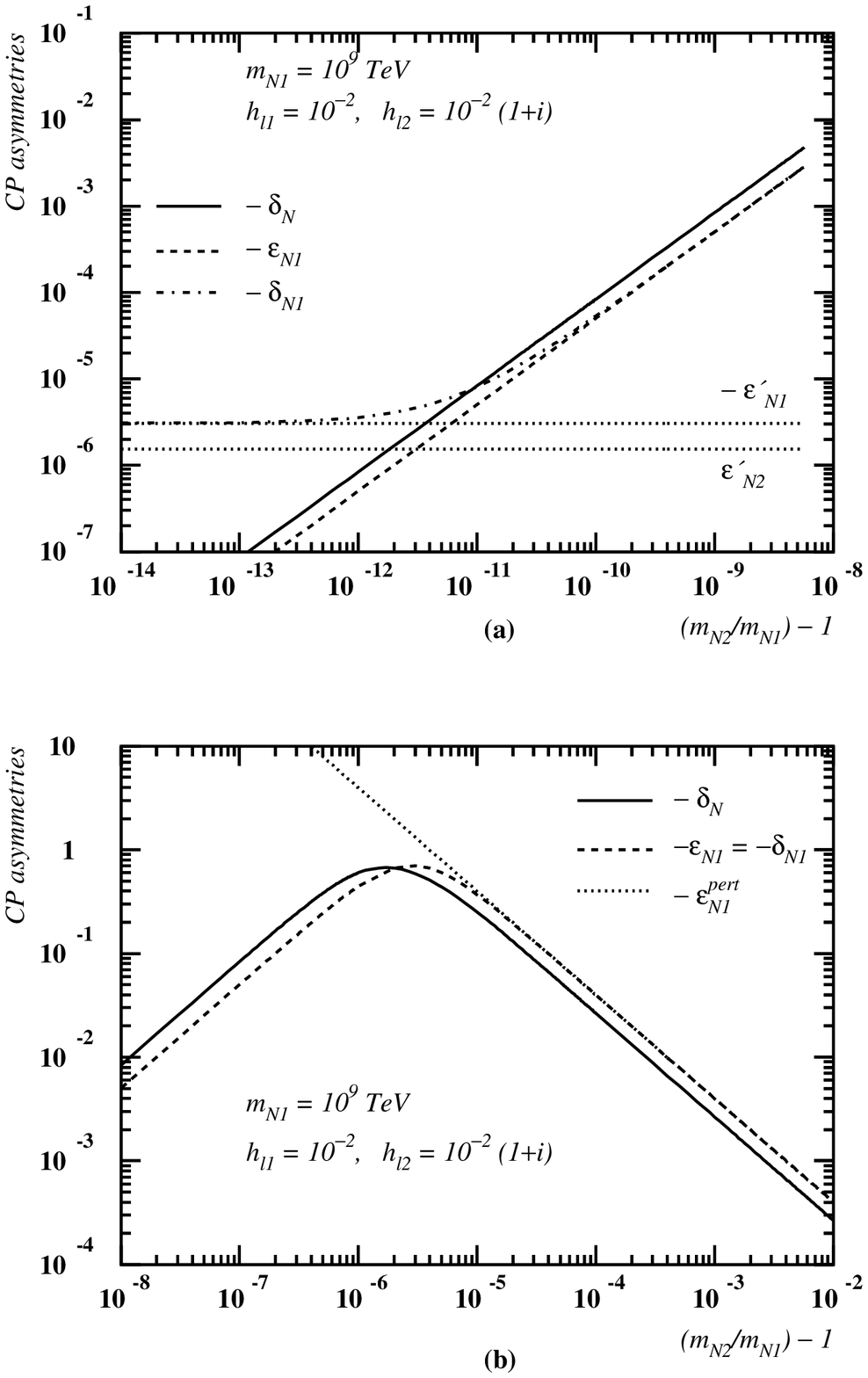}
 \end{center}

\fcaption{Numerical estimates of CP asymmetries versus
$m_{N_2}/m_{N_1} -1 $ in scenario II.}\label{fig:5}

\end{figure}

\begin{figure}[hb]
   \leavevmode
 \begin{center}
   \epsfxsize=11.cm
   \epsffile[0 0 425 425]{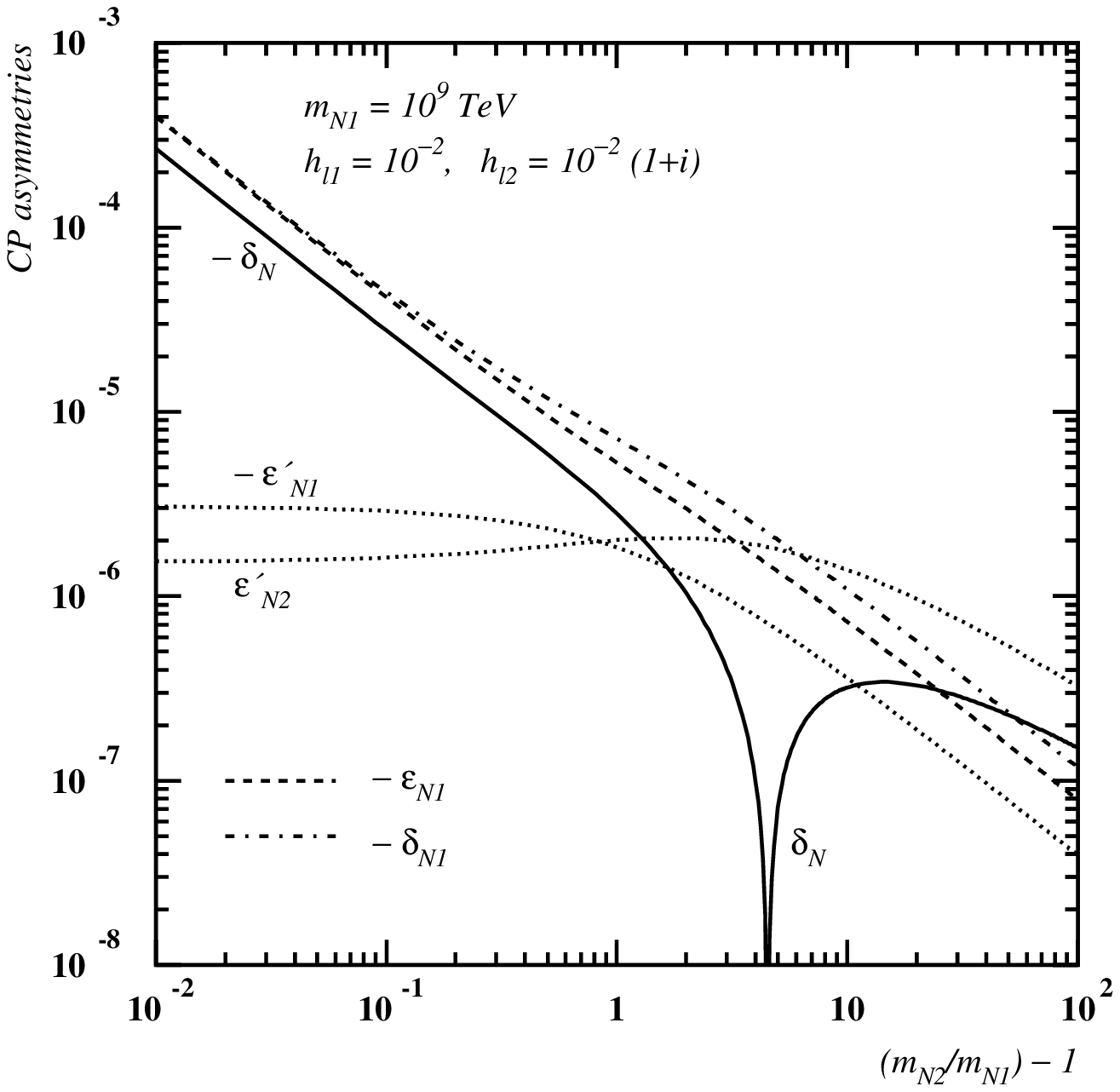}
 \end{center}

\fcaption{Numerical estimates of CP asymmetries as a function
of $m_{N_2}/m_{N_1} -1 $ in scenario II.}\label{fig:6}

\end{figure}

To gauge better the dependence of CP asymmetries on the heavy neutrino
masses, we  shall  adopt two simple  scenarios with   two-right handed
neutrinos that mix actively with one lepton family $l$ only:
\begin{eqnarray}
\label{scenario}
\mbox{I.} && m_{N_1}\, =\, 10\ \mbox{TeV}\, ,\qquad h_{l1}=10^{-6},\
\quad h_{l2}=10^{-6}(1+i)\, ,\nonumber\\
\mbox{II.} && m_{N_1}\, =\, 10^9\ \mbox{TeV}\, ,\qquad h_{l1}=10^{-2},\
\quad h_{l2}=10^{-2}(1+i)\, .
\end{eqnarray}
We assume that $N_2$ is  always heavier than $N_1$, i.e.\ $m_{N_1}\leq
m_{N_2}$.    The  above   two   scenarios comply    qualitatively with
Sakharov's third requirement of out-of-equilibrium condition (see also
discussion in Section 8).  In view of Eq.\ (\ref{dCP}), both scenarios
I and  II given above  represent maximal  cases  of CP violation  with
$\delta_{CP}=1$.  Therefore, results for any  other model may  readily
be  read off  by multiplying the   CP asymmetries with the appropriate
model-dependent factor $\delta_{CP}$.

Figure \ref{fig:4} exhibits the dependence of the  CP asymmetries as a
function of  the parameter $x_N$ for  scenario I.  The parameter $x_N$
defined in Eq.\ (\ref{xN}) is a measure of mass degeneracy for the two
heavy Majorana  neutrinos  $N_1$ and $N_2$.  We   divide  the range of
values for the parameter  $x_N$ into two  regions: the first region is
plotted in Fig.\ \ref{fig:4}(a) and  pertains to the kinematic  domain
where resonant CP violation due to  heavy-neutrino mixing occurs.  The
second one, shown  in Fig.\  \ref{fig:4}(b), represents the  kinematic
range far away from  the resonant CP-violating phenomenon.  The dotted
line   in    Fig.\    \ref{fig:4}(a) gives       the  prediction    of
$\varepsilon_{N_1}$,  when  Eq.\ (\ref{epsN1})  is   calculated in the
conventional    finite-order  perturbation  theory.         Obviously,
$\varepsilon^{pert}_{N_1}$ diverges  for sufficiently small  values of
$x_N$, e.g.\   $x_N  < 10^{-13}$.   If   resummation  of the  relevant
fermionic self-energy   graphs is    considered, the  prediction   for
$\varepsilon_{N_1}$ becomes  analytic and is  given by the dashed line
in  Fig.\ \ref{fig:4}.  The   $\varepsilon_{N_1}$ line shows a maximum
for $x_N\approx 10^{-13}$.  In agreement with  the conditions in Eqs.\ 
(\ref{CPcond}) and (\ref{dCP}), CP  violation may resonantly  increase
up to order unity.\cite{APRL,ANPB} The solid line in Fig.\ \ref{fig:4}
displays  the dependence  of the CP-violating  parameter $\delta_N$ in
Eq.\ (\ref{deltaN})  on $x_N$, where $\varepsilon'$-type contributions
are  included.  The latter are very  small in this  scenario, so as to
account for   the  BAU, e.g.\  $\varepsilon'_{N_1}\approx  10^{-16}$.  
Finally,  we comment  on the   fact  that $\delta_N$ vanishes in   the
CP-invariant limit $x_N\to  0$, as  it should be  on account  of  Eq.\ 
(\ref{CPinv}).

Figures \ref{fig:5} and \ref{fig:6}   give numerical estimates  of  CP
asymmetries in scenario II. The difference of  this model with that of
scenario I  is   that the  $\varepsilon'$-type   effects may  not   be
negligible in  the off-resonant region,  as  can be seen  from  Figs.\ 
\ref{fig:5}(a)  and  \ref{fig:6}.  In  particular,  for values  of the
parameter  $x_N  <   10^{-11}$    or   $x_N >    1$,   the  individual
$\varepsilon'_{N_1}$- and $\varepsilon'_{N_2}$-type contributions  may
prevail over  the $\varepsilon$-type ones.   Models with $x_N> 1$ have
been    extensively       discussed                 in             the
literature.\cite{FY,MAL,CEV,epsilonprime} Numerical estimates for such
models  are displayed  in Fig.\    \ref{fig:6}.   We first  focus  our
attention on the domain with  $x_N<10^{-2}$.  In Fig.\ \ref{fig:5}(a),
we  observe     that   $\varepsilon'_{N_1}$ and  $\varepsilon'_{N_2}$,
represented by the  dotted lines,  do not  vanish in the  CP-invariant
limit     $x_N\to 0$,  as   opposed   to  $\varepsilon_{N_1}$.   As  a
consequence, the CP  asymmetry $\delta_{N_1}$ in Eq.\ (\ref{deltaNi}),
in which both $\varepsilon_{N_1}$- and $\varepsilon'_{N_1}$-type terms
are considered  within  our formalism,  does  not  vanish either.  The
reason  is that  the physical CP-violating   parameter in this  highly
degenerate mass  regime    for $N_1$  and   $N_2$  is  the  observable
$\delta_N$ defined in  Eq.\  (\ref{deltaN}).  In fact, $\delta_N$  and
$\varepsilon_{N_1}$   share   the   common  feature  that  both   tend
consistently to  zero as $x_N\to 0$.  This  fact must be considered to
be  one of the   successes of  the  resummation  approach.  Again,  CP
violation is  resonantly    amplified,  when the condition  in    Eq.\ 
(\ref{CPcond}) is satisfied, as can be seen from Fig.\ \ref{fig:5}(b).
Finally, we must  remark that  $-\delta_N$  flips sign  and eventually
becomes negative   for   $x_N\gg 1$,    as can   be   seen from  Fig.\ 
\ref{fig:6}.  However,  in this  kinematic  range, we must  consider a
further refinement into the definition  of $\delta_N$.  The effect  of
the different  dissipative  Boltzmann  factors multiplying   the decay
rates  of the heavy Majorana  neutrinos  $N_1$ and $N_2$  must also be
included in $\delta_N$.  These phenomena will be taken into account in
Section 8.

\setcounter{section}{7}
\setcounter{equation}{0}
\section{Unitarity and CPT invariance in the resummation approach}\label{sec:7}
\noindent
It is interesting to see  how the  resummation approach preserves  CPT
invariance and   unitarity.\cite{ANPB}  An  immediate   consequence of
unitarity and CPT symmetry is  that CP violation in the  $L$-violating
scattering process  $L\Phi^\dagger  \to  L^C\Phi$ is  absent  to order
$h^6_{li}$.\cite{RCV,BP} We will  concentrate on the resonant part  of
the amplitude, as it is the dominant one.

Our aim is to show that to `one loop',
\begin{equation}
  \label{DCP}
\Delta_{\rm CP}\ =\ \int d{\rm LIPS}\ 
|{\cal T}^{\rm res}_{L\Phi^\dagger \to  L^C\Phi}|^2\
-\ \int d{\rm LIPS}\
|\overline{\cal T}^{\rm res}_{L^C\Phi \to  L\Phi^\dagger}|^2\ =\ 0,
\end{equation}
where LIPS stands for the two-body Lorentz-invariant phase space.  For
simplicity,  we omit external  spinors and absorb irrelevant constants
in the definition of the Yukawa-coupling matrix $h = (h_{l1},h_{l2})$.
Using matrix notation, the resummed transition amplitudes are written
\begin{equation}
  \label{Tres}
{\cal T}^{\rm res}_{L\Phi^\dagger \to  L^C\Phi}\ =\ 
hP_R\, S(\not\! p)\, P_R h^T\, ,\qquad
\overline{\cal T}^{\rm res}_{L^C\Phi \to  L\Phi^\dagger}\ =\ 
h^*P_L\, \bar{S}(\not\! p)\,P_L h^\dagger\, ,
\end{equation}
with    $\bar{S}(\not\!\!   p)    =   S^T(\not\!\!   p)$   being   the
CP/T-conjugate propagator matrix of $S (\not\!\!  p)$.  In writing the
CP/T-conjugate amplitude  $\overline{\cal   T}^{\rm res}_{L^C\Phi  \to
  L\Phi^\dagger}$, we have  employed  the identities  (\ref{idCP}) for
spinor   objects and  made use   of  the  rotational symmetry of   the
amplitude.   The  latter has  the   effect  of  reversing the  spatial
components of  the four momenta.  We  also neglect possible P-odd spin
correlations involving  external leptons since  they  will be averaged
away when forming the matrix element squared.

We   start the  proof  by noticing   that   as  a consequence of   CPT
invariance,
\begin{equation}
  \label{CPT}
|{\cal T}^{\rm res}_{L\Phi^\dagger \to  L\Phi^\dagger}|^2\ =\ 
|{\cal T}^{\rm res}_{L^C\Phi \to  L^C \Phi}|^2\, .
\end{equation}
This equality is indeed valid, since 
\begin{equation}
|hP_R\, S(\not\! p)\, P_Lh^\dagger|\ =\ |h^*P_L\, S^T(\not\! p)\, P_R h^T|\
=\ |h^*P_L\, \bar{S}(\not\! p)\, P_R h^T|\ .
\end{equation}

Unitarity of the   theory prescribes the following  relation governing
the resummed propagators:
\begin{equation}
  \label{OT}
S^{-1}(\not\! p)\ -\ S^{-1\dagger}(\not\! p) \ =\ -i \int
d{\rm LIPS} \not\! p (h^T h^* P_L\,  +\,  h^\dagger h P_R)\ .
\end{equation}
This last relation is also known as the  optical theorem. Based on the
optical theorem, we can prove the equality
\begin{equation}
  \label{Uni}
\int d{\rm LIPS}\
|{\cal T}^{\rm res}_{L\Phi^\dagger \to L\Phi^\dagger, L^C\Phi}|^2\ 
=\ \int d{\rm LIPS}\
|\overline{\cal T}^{\rm res}_{L^C\Phi \to  L\Phi^\dagger,
L^C\Phi}|^2\ .
\end{equation}
Indeed, using Eq.\ (\ref{OT}), we find 
\begin{eqnarray}
  \label{Tres1}
\int d{\rm LIPS}\
|{\cal T}^{\rm res}_{L\Phi^\dagger \to L\Phi^\dagger, L^C\Phi}|^2
&& \nonumber\\
&&\hspace{-4cm}= \int d{\rm LIPS}\
hP_R\, S(\not\! p)\,  \not\! p (h^T h^* P_L\,  +\,  h^\dagger h P_R)\,
S^\dagger (\not\! p)\, P_L h^\dagger\nonumber\\
&&\hspace{-4cm}= -i\, hP_R\, [S(\not\! p)\, -\, S^\dagger (\not\! p)]\, 
P_L h^\dagger\ =\ 2\, hP_R\, {\rm Im} S(\not\! p)\, P_Lh^\dagger\, ,
\end{eqnarray}
and for the CP-conjugate total rate, 
\begin{eqnarray}
  \label{Tres2}
\int d{\rm LIPS}\
|{\cal T}^{\rm res}_{L^C\Phi \to L\Phi^\dagger, L^C\Phi}|^2
& =&   2\, h^*P_L\, {\rm Im} \bar{S}(\not\! p)\, P_Rh^T\, \nonumber\\
&=& 2\, hP_R\, {\rm Im} S(\not\! p)\, P_Lh^\dagger\, .
\end{eqnarray}
As the RHSs of  Eqs.\ (\ref{Tres1}) and  (\ref{Tres2}) are equal,  the
equality (\ref{Uni}) is  obvious.  Subtracting  Eq.\ (\ref{CPT})  from
Eq.\ (\ref{Uni}),  it is not  difficult to show that $\Delta_{\rm CP}$
vanishes at the one-loop  resummed level.  We  should remark that  the
resummation  approach\cite{ANPB}  satisfies   CPT  and unitarity  {\em
  exactly},   without recourse to  any   re-expansion of the  resummed
propagator.  If we also include  resummed amplitudes subleading in the
Yukawa  couplings, then residual  CP-violating terms that are formally
of   order $h^8_{li}$ and  higher occur  in  $\Delta_{\rm CP}$.  These
terms  result from   the   interference of  two  resummed   amplitudes
containing one-loop vertex graphs.  Because of unitarity, however, the
residual CP-violating terms of order $h^8_{li}$ and $h^{10}_{li}$ will
cancel at two loops together with respective CP-violating terms coming
from one-loop $2\to 4$ scatterings, and so on.

In   the   approach   under   consideration,\cite{ANPB}   the physical
transition amplitude    is obtained  by    sandwiching  the   resummed
propagators  between matrix  elements  related  to  initial and  final
states   of the resonant process.   Therefore,   diagonalization of $S
(\not\!   p)$  is  no longer   necessary,  thereby  avoiding  possible
singularities   emanating  from    non-diagonalizable    (Jordan-like)
effective  Hamiltonians      [or   equivalently $S^{-1}     (\not\!\!  
p)$].\cite{ANPB} In    fact,  such effective    Hamiltonians represent
situations in which the CP-violating  mixing between the two  unstable
particles reaches its      maximum and physical  CP   asymmetries  are
therefore large.  In such a case, the  complex mass eigenvalues of the
effective Hamiltonian are exactly equal.

To  see this  point  in  more detail, let   us consider  the following
effective Hamiltonian for the $N_1N_2$ system:
\begin{equation}
\label{effHam}
{\cal H}(\not\! p)\ =\ \left[ \begin{array}{cc}
m_1 - \widehat{\Sigma}_{11}(\not\! p) & -\widehat{\Sigma}_{12}(\not\! p)\\
-\widehat{\Sigma}_{21}(\not\! p) & m_2 - \widehat{\Sigma}_{22}(\not\! p)
\end{array} \right]\ \approx\
\left[ \begin{array}{cc}
m_N + a - i|b|  & -ib\\
-ib^* & m_N - a - i|b| \end{array} \right],
\end{equation}
in the approximation $\not\!\! p \to  m_N \approx m_1\approx m_2$.  In
Eq.\ (\ref{effHam}), the parameters $a$  and $b$ are real and complex,
respectively, and $m_1 = m_N+a$,  $m_2 = m_N-a$. The complex parameter
$b$ represents    the    absorptive part   of   the  one-loop neutrino
transitions $N_i\to N_j$.  Unitarity requires  that the determinant of
the absorptive part  of ${\cal H}(\not\!\!  p)$  be non-negative.  For
the      effective   Hamiltonian  (\ref{effHam}),  the   corresponding
determinant is zero.  One-generation models  naturally lead to such an
absorptive effective  Hamiltonian. If $a =  |b|$, the two complex mass
eigenvalues of ${\cal  H} (\not\!\!  p)$   are exactly degenerate  and
equal   to $m_N-i|b|$.  Then,   the   effective Hamiltonian cannot  be
diagonalized via a similarity transformation  in this limit, i.e.\ the
respective diagonalization matrices become singular.\cite{ANPB}

An interesting  question   one may  raise   in  this  context  is  the
following. Since models with non-diagonalizable effective Hamiltonians
lead to an exact equality between their  complex mass eigenvalues, how
then   can this fact be  reconciled  with  the CP-invariance condition
(\ref{CPinv})?  According  to the condition  (\ref{CPinv}), any effect
of CP violation must vanish identically, and should not even be large!
To resolve this  paradox, one should notice  that in the presence of a
large particle mixing, the mass eigenstates  of $S^{-1} (\not\!\!  p)$
are   generally    non-unitary    among themselves,      whereas   the
OS-renormalized mass  eigenstates\cite{KP} form a well-defined unitary
basis (or      any other  renormalization     scheme that    preserves
orthonormality of  the Hilbert space),  upon which perturbation theory
can be formulated order by  order.  Therefore, the field-theoretic  OS
renormalized masses  are    those that enter    the  condition of   CP
invariance   given by  Eq.\ (\ref{CPinv}).   Consequently,  if the two
complex mass eigenvalues of the effective  Hamiltonian are equal, this
does not  necessarily entail an equality  between  their respective OS
renormalized masses, and   therefore   absence  of CP   violation   as
well.\cite{ANPB}

\setcounter{section}{8}
\setcounter{equation}{0}
\section{Boltzmann equations}\label{sec:8}
\noindent
The thermodynamic evolution  of the system in  the radiation-dominated
era of the  Universe may be described  by a  set of coupled  Boltzmann
equations  (BE's).\cite{KW,EWK&SW,KT,HT} These equations determine the
time evolution of the  lepton-number asymmetry which will be converted
into the   observed BAU   by sphalerons.   We  shall solve    the BE's
numerically and present  results for the expected  BAU  within the two
different democratic-type scenarios  I and II discussed  in Section 6. 
Finally, we will give estimates of the finite-temperature effects, and
discuss their impact on resonant CP violation through mixing.

Before solving  numerically the  BE's,  it  is instructive  to discuss
first the   out-of equilibrium constraints on   heavy neutrino decays. 
Sakharov's third necessary condition  requires that the expansion rate
of the Universe be smaller  than the decay  rate of any  $L$-violating
process.  The most dominant $L$-violating process  is the decay of the
heavy  Majorana  neutrinos   themselves,   $\Gamma_{N_i}$  (cf.\  Eq.\ 
(\ref{GammaN})).  To  a  good approximation,  we have  the approximate
inequality
\begin{equation}
\label{Sakh3}
\Gamma_{N_i} (T=m_{N_i})\ \stackrel{\displaystyle <}{\sim}\
2\, K\, H(T=m_{N_i})\, ,
\end{equation}
where $K\approx 1$--$1000$ is   a factor quantifying the deviation  of
the decay rates from the expansion rate of the Universe, and $H(T)$ is
the Hubble parameter
\begin{equation}
\label{Hubble}
H(T)\ =\ 1.73\, g_*^{1/2}\, \frac{T^2}{M_{\rm Planck}}\ .
\end{equation}
with $M_{\rm   Planck}  =  1.2\times   10^{19}$  GeV  and  $g_*\approx
100$--400 being   the number of   active degrees  of freedom in  usual
extensions of the SM.  Then, the out-of equilibrium constraint in Eq.\ 
(\ref{Sakh3}) translates into the bound
\begin{equation}
\label{hli_bound}
|h_{li}|^2\ \stackrel{\displaystyle <}{\sim}\
7.2\,K \times 10^{-14}\, \Big( \frac{m_{N_i}}{1\ \mbox{TeV}}\Big)\, .
\end{equation}
Although  not mandatory, this very  last constraint  may be applied to
all Yukawa couplings.

As we have discussed in Section 2 (see also Eq.\ (\ref{BLrel})), above
the  electroweak   phase transition  the  $B+L$-sphaleron interactions
convert approximately one-third of the lepton-to-entropy density ratio
$Y_L  = n_L/s$ into a baryon-to-entropy  density  ratio $Y_B = n_B/s$,
i.e.\ \cite{BS,HT}
\begin{equation}
\label{YB_YL}
Y_B\approx -\, \frac{1}{3}\, Y_L\, \approx -\,\frac{1}{3K}\
\frac{\delta_{N_i}}{g_*}\ .
\end{equation}
The  last approximate  equality represents the  asymptotic solution of
the relevant BE's.\cite{MAL,APRD}  {}From Eq.\  (\ref{YB_YL}), we  see
that  $Y_B$ can   be in the  observed   ball park, i.e.\   $Y_B\approx
10^{-10}$,  if  $|\delta_{N_i}|/K$ are of  order $10^{-7}$--$10^{-6}$. 
Clearly, CP asymmetries of order unity  allow for very large values of
$K$.  As a  consequence, the thermal plasma can  then be rather  dense
and  the conditions of kinetic  equilibrium in BE's can comfortably be
satisfied even within the minimal leptogenesis scenario under study.

We now turn  to the discussion of BE's.    The lepton asymmetry  for a
system with two heavy Majorana neutrinos  is determined by the coupled
system of BE's\cite{KT,MAL}
\begin{eqnarray}
\label{BENi}
\frac{dn_{N_i}}{dt}\, +\, 3Hn_{N_i} &=& 
-\, \Big( \frac{n_{N_i}}{n^{eq}_{N_i}}\, -\, 1\Big)\, \gamma_{N_i}\\ 
\label{BElept}
\frac{dn_L}{dt}\, +\, 3Hn_{L} &=& \sum\limits_{i=1}^2\, \Big[\, \delta_{N_i}
\, \Big( \frac{n_{N_i}}{n^{eq}_{N_i}}\, -\, 1\Big)\, -\, \frac{n_L}{2n^{eq}_l}
\, \Big]\, \gamma_{N_i}\ -\
\frac{n_L}{n^{eq}_l}\, \gamma_{\sigma}\, ,
\end{eqnarray}
where $n_{N_i}$, $n_L=n_l-n_{\bar{l}}$ are the densities of the number
of   $N_i$  and  the  lepton-number    asymmetry,   respectively,  and
$n^{eq}_{N_i}$ and $n^{eq}_l$ are their values in thermal equilibrium.
The  Hubble parameter $H=(dR/dt)/R$  determines  the expansion rate of
the Universe  and  also depends  on the  temperature  $T$, through the
relation  in    Eq.\  (\ref{Hubble}).    In   Eqs.\  (\ref{BENi})  and
(\ref{BElept}), $\gamma_{N_i}$ and $\gamma_{\sigma}$ are the decay and
scattering collision terms, respectively:
\begin{eqnarray}
\label{gNi}
\gamma_{N_i}& =& n^{eq}_{N_i}\, \frac{K_1 (m^2_{N_i}/T)}{K_2(m^2_{N_i}/T)}\,
\Gamma_{N_i}\, ,\\
\label{gsigma}
\gamma_\sigma &=& \frac{T}{8\pi^4}\, \int_{s_{\rm thr}}^\infty\, 
ds\, s^{3/2}\, K_1 (\sqrt{s}/T)\, \sigma'(s)\, .
\end{eqnarray}
Here, $s_{\rm thr}$ is  the threshold  of a  generic process  $a+b \to
c+d$, and
\begin{equation}
  \label{sigmapr}
\sigma'(s)\ =\ \frac{1}{4}\ \theta(\sqrt{s} - m_a - m_b)\ 
\lambda\Big( 1,\ \frac{m^2_a}{s},\ \frac{m^2_b}{s}\,\Big)\ 
\hat{\sigma} (s)  
\end{equation}
with $\lambda (x,y,z)   =  (x-y-z)^2 -  4yz$.   In Eq.\   (\ref{gNi}),
$K_1(z)$ and $K_2(z)$ are  the  modified Bessel functions  defined  in
Ref.\cite{MA&IAS} The cross section $\hat{\sigma}(s)$ mainly comprises
the scatterings   $L^C\Phi\to  L\Phi^\dagger$  and   its  CP-conjugate
process  $L\Phi^\dagger\to  L^C\Phi$, and  is  evaluated  at $T=0$  by
subtracting all     those real intermediate  contributions   that have
already been taken  into account in the  direct and inverse decays  of
heavy     Majorana     neutrinos.\cite{EWK&SW}   The  collision   term
$\gamma_{\sigma}$ acts  as a  CP-conserving  depletion term,  which is
formally   of order $\gamma^2_{N_i}$  at  $T\approx m_{N_i}$.\cite{KT}
There  is  also the  $\Delta  L =2$ reaction $\Phi\Phi \leftrightarrow
LL$, which    is much   weaker   than  the latter, as    long   as the
out-of-equilibrium  constraint    on the  Yukawa    couplings in  Eq.\ 
(\ref{hli_bound})  is  imposed.    Finally,   there exist   additional
contributions  to the  BE's,\cite{MAL}  coming from processes such  as
$N_i L  \leftrightarrow   Q_i \bar{t}_R$, $N_i  Q_i  \leftrightarrow L
t_R$.  These contributions are quite strong at very high temperatures,
$T  \gg m_{N_i}$,  and lead to   a decoherence  phenomenon between the
heavy neutrinos $N_1$ and $N_2$.  At  the crucial leptogenesis  epoch,
when  $T\approx  m_{N_i}$, the rates    of  the latter  processes  are
kinematically suppressed and smaller than the decay rates of the heavy
Majorana neutrinos.\cite{KT}

Many  applicable assumptions are   involved  in BE's  (\ref{BENi}) and
(\ref{BElept}).  More details may  be found in Ref.\cite{KT} First, we
have  considered   the    Friedmann-Robertson-Walker  model    in  the
non-relativistic limit.  Second, we have adopted the Maxwell-Boltzmann
statistics, which is  a good approximation  in the absence of  effects
that  originate from Bose  condensates or  arise  from a degeneracy of
many Fermi degrees of freedom.  Third, we have assumed that the lepton
and Higgs weak isodoublets, $L$ and $\Phi$, are practically in thermal
equilibrium,  and    neglected  high   orders in   $n_L/n^{eq}_l$  and
$\delta_{N_i}$.  In  this context, it  has also  been assumed that the
different particle species are in kinetic  equilibrium, i.e.\ that the
particles may rapidly    change their kinetic   energy through elastic
scatterings but the processes  responsible for a  change of the number
of  particles  are  out   of equilibrium.     These out-of-equilibrium
reactions are described by the BE's (\ref{BENi}) and (\ref{BElept}).

To solve these BE's numerically, it proves useful  to make the following
change of variables:
\begin{equation}
x\ =\ \frac{m_{N_1}}{T}\ ,\qquad t\ =\ \frac{1}{2H(T)}\ =\ 
\frac{x^2}{2H(x=1)}\ .
\end{equation}
Such an ansatz is also valid for the  radiation-dominated phase of the
Universe    while  baryogenesis takes  place.    Then,  we  define the
parameters
\begin{equation}
\label{Kparam}
K\ =\ \frac{K_1(x)}{K_2(x)}\, \frac{\Gamma_{N_1}}{H(x=1)}\ ,
\qquad \gamma\ =\ \frac{K_2(x)K_1(\xi x)}{K_1(x)K_2(\xi x)}\, 
\frac{\Gamma_{N_2}}{\Gamma_{N_1}}\ ,
\end{equation}
with  $\xi = m_{N_2}/m_{N_1}\ge  1$.   In addition,  we  introduce the
quantities $Y_{N_i}  = n_{N_i}/s$ and $Y_L =  n_L/s$, where $s$ is the
entropy density.  In  an isentropically expanded Universe, the entropy
density has  the time dependence  $s(t)=\mbox{const.}\times R^{-3}(t)$
and may  be related to the  number density of  photons, $n_\gamma$, as
$s=g_*  n_\gamma$, where  $g_*$ is  given after  Eq.\  (\ref{Hubble}). 
Emplyoing the above definitions and relations among the parameters, we
obtain the BE's  in terms of the  new quantities $Y_{N_1}$,  $Y_{N_2}$
and $Y_L$:
\begin{eqnarray}
\label{BEYN1}
\frac{dY_{N_1}}{dx} &=& -\, (Y_{N_1} - Y^{eq}_{N_1}) Kx^2\, ,\\
\label{BEYN2}
\frac{dY_{N_2}}{dx} &=& -\, (Y_{N_2} - Y^{eq}_{N_2}) \gamma Kx^2\, ,\\
\label{BEYL}
\frac{dY_L}{dx} &=& \Big[\, (Y_{N_1}-Y^{eq}_{N_1})\delta_{N_1}\, +\, 
(Y_{N_2} - Y^{eq}_{N_2} )\gamma\delta_{N_2}\, -\, \frac 12 g_* Y_L
(Y^{eq}_{N_1}+\gamma Y^{eq}_{N_2})  \nonumber\\
&&-\, g_* Y_L Y^{eq}_{N_1} \frac{\gamma_\sigma}{\gamma_{N_1}}\, 
\Big]\, Kx^2\, .
\end{eqnarray}
The heavy-neutrino    number-to-entropy      densities in  equilibrium
$Y^{eq}_{N_i}(x)$ are given by\cite{EWK&SW}
\begin{equation}
\label{YeqN}
Y^{eq}_{N_1}(x)\ =\ \frac{3}{8g_*}\, \int_{x}^\infty\, 
dz\, z\, \sqrt{z^2-x^2}\, e^{-z}\ =\ \frac{3}{8g_*}\, x^2\, K_2(x)\, , 
\end{equation}
and $Y^{eq}_{N_2}(x)=Y^{eq}_{N_1}(\xi x)$. The differential  equations
(\ref{BEYN1})--(\ref{BEYL})  are solved numerically, using the initial
conditions
\begin{equation}
\label{InBE}
Y_{N_1}(0)\ =\ Y_{N_2}(0)\ =\  Y^{eq}_{N_1}(0)\ =\ Y^{eq}_{N_2}(0)
\quad \mbox{and}\quad Y_L(0)=0\, .
\end{equation}
These initial conditions merely  reflect  the fact that  our  Universe
starts evolving   from a lepton-symmetric  state, in   which the heavy
Majorana neutrinos are   originally in thermal  equilibrium.  Here, we
should   remark that  the   low-temperature  limit  of the   numerical
predictions   does  not strongly  depend  on   the  initial conditions
(\ref{InBE}), if $L$-violating interactions are in thermal equilibrium
at $T\gg m_{N_i}$. The reason  is that at  very high temperatures, the
BE's (\ref{BEYN1})--(\ref{BEYL})  exhibit  a running independent fixed
point, and  any initial value  for  $Y_{N_1}$, $Y_{N_2}$ and  $Y_L$ is
rapidly  driven to   the  thermal-equilibrium  values  given by   Eq.\ 
(\ref{InBE}).\cite{EAP}  After   the  evolution of  the   Universe  to
temperatures  much  below $m_{N_1}$, a   net lepton asymmetry has been
created. This lepton asymmetry will then be converted into the BAU via
the sphalerons.  During a  first  order electroweak phase  transition,
the produced excess in $L$ is also encoded as  an excess in $B$, which
is  given by   Eq.\  (\ref{BLrel}).\cite{BS,HT} The  observed   BAU is
$Y^{obs}_B = (0.6 -  1)\times 10^{-10}$,\cite{KT} which corresponds to
an excess  of leptons $-Y^{obs}_L  \approx 10^{-9} - 10^{-10}$. In the
latter estimate, we  have included the  possibility of  generating the
BAU via an individual lepton asymmetry.\cite{Dreiner/Ross}

\begin{figure}[ht]
   \leavevmode
 \begin{center}
   \epsfxsize=11.cm
   \epsffile[0 0 539 652]{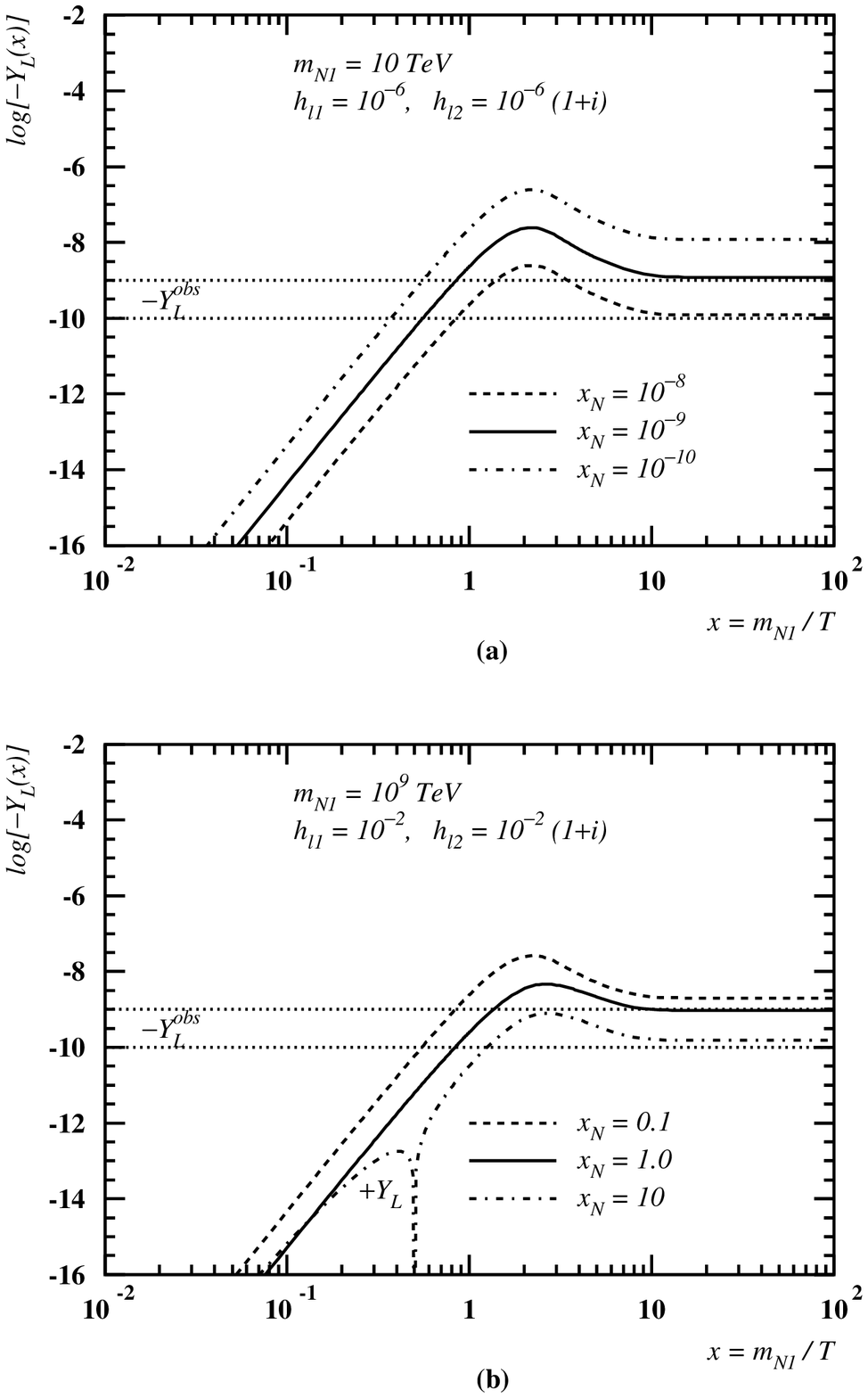}
 \end{center}
 
 \fcaption{Lepton asymmetries for selected heavy Majorana neutrino
   scenarios.}\label{fig:7}

\end{figure}

Figure \ref{fig:7} shows the dependence  of $Y_L(x)$ on  $x=m_{N_1}/T$
for two representative scenarios  defined in Eq.\ (\ref{scenario}) for
different  values  of   $x_N = m_{N_2}/m_{N_1}    - 1$.\cite{APRD} The
observed range for $Y_L$, $Y^{obs}_L$, is indicated with two confining
horizontal  dotted lines.   In  scenario I  (Fig.\  \ref{fig:7}(a)), a
heavy-neutrino mass splitting  $x_N$ of order  $10^{-9}$--$10^{-8}$ is
sufficient to  account   for the BAU.  For  comparison,    it is worth
mentioning that the degree of mass  degeneracy between $K_L$ and $K_S$
is  of   order $10^{-15}$,  which  is  by  far  smaller  than  the one
considered here.  We find  that the $\varepsilon$-type CP violation is
dominant, whereas $\varepsilon'$-type effect are extremely suppressed.
Numerical  estimates for  the second scenario  are  displayed in Fig.\ 
\ref{fig:7}(b).   This scenario   is  closer to   the traditional  one
considered in Ref.\cite{FY} Here,  it is not  necessary to have a high
degree  of degeneracy  for   $N_1$ and  $N_2$   to  get  sufficient CP
violation for    the  BAU.  In  this   case, both  $\varepsilon$-  and
$\varepsilon'$-type mechanisms of CP violation  are equally important. 
Therefore, the main consequence of  $\varepsilon$-type CP violation is
that the  leptogenesis scale may  be as low as 1  TeV, even for models
with universal Yukawa couplings.\cite{APRD}

In the scenario  of leptogenesis induced  by mixing of heavy  Majorana
neutrinos, one may have to worry about effects, which could affect the
resonant  condition of   CP  violation  in Eq.\  (\ref{CPcond}).   For
instance, there may be broadening  effects at high temperatures due to
collisions among particles.    Such effects will  contribute terms  of
order   $h^4_{li}$    to   the  $N_i$ widths      and   are small   in
general.\cite{EWK&SW,Roulet}  On  the other  hand,  finite temperature
effects on the $T=0$ masses of particles may  be significant.  Because
of the SM gauge interactions, the leptons and the Higgs fields receive
appreciable thermal masses,\cite{HAW,MEC,CKO} i.e.
\begin{eqnarray}
\label{thermal}
\frac{m^2_L(T)}{T^2} &=& \frac{1}{32}\, (3g^2\, +\, g'^2)\ \approx\
0.044\, ,
\end{eqnarray}
where $g$ and $g'$ are  the SU(2)$_L$ and  U(1)$_Y$ gauge couplings at
the running scale $M_Z$.  The  isosinglet heavy neutrinos also acquire
thermal masses through Yukawa interactions \cite{HAW}, i.e.
\begin{equation}
\label{mN(T)}
\frac{m^2_{N_i}(T)\, -\, m^2_{N_i}(0)}{T^2}\ =\ \frac{1}{16}\, |h_{li}|^2\, .
\end{equation}
Such a $T$-dependent  mass shift is  small and comparable to the $N_i$
widths at $T\approx  m_{N_i}$.  Therefore, it is easy  to see that the
condition  for resonant   CP     violation through mixing    in   Eq.\ 
(\ref{CPcond}) is  qualitatively satisfied.  Finally, the  Higgs field
also  receives   appreciable thermal contributions.    The  authors of
Ref.\cite{CKO}  have  computed the one-loop   Higgs  thermal mass, and
found  that $M_\Phi  (T)/T  \stackrel{\displaystyle  <}{\sim} 0.6$ for
values of the  Higgs-boson mass favoured  by LEP2, i.e.\  $M_H < 200$. 
In  this range of Higgs masses,  the  thermal widths $\Gamma_{N_i}(T)$
will be reduced with respect to $\Gamma_{N_i}(0)$ by  a factor of 2 or
3  due   to sizeable    phase-space  corrections.  Nevertheless,   the
influence on the resonant phenomenon of CP violation through mixing is
not dramatic when the latter effects are included, and therefore large
leptonic CP asymmetries are still conceivable.

\newpage

\setcounter{section}{9}
\setcounter{equation}{0}
\section{Low-energy phenomenology of heavy Majorana neutrinos}\label{sec:9}
\noindent
Whether heavy Majorana neutrinos can lead to interesting phenomenology
at collider and  lower energies is an  issue that strongly  depends on
the out-of-equilibrium constraint  given by Eq.\ (\ref{hli_bound}). If
this   constraint is applied  to  all  lepton families, heavy Majorana
neutrinos have very  little impact on collider phenomenology. However,
this very last statement is  rather model dependent.  One can imagine,
for example,  a  scenario in  which  $\Delta L_e$-violating  operators
exist and are out-of-equilibrium, and the $\mu$  and $\tau$ sectors do
not communicate any interaction to the  electron sector, i.e.\ $\Delta
(L_e - L_\mu) = \Delta (L_e -  L_\tau) = 0$. Since sphalerons conserve
the individual quantum  number $B/3 - L_e$ (see  also Section 2),  the
observed  baryonic asymmetry can be preserved   in an excess of $L_e$,
independently     of   whether    $\Delta  L_\mu$-   and/or    $\Delta
L_\tau$-non-conserving operators are in thermal equilibrium or not. As
we will discuss below,  such scenarios with  strong mixing in the muon
and tau  sectors  only can  give rise   to a  variety  of  new-physics
phenomena in a strength that can be probed in laboratory experiments.

\subsection{Lepton-flavour and/or number processes}
\noindent  
Heavy Majorana neutrinos with masses in the range 0.2  -- 1 TeV may be
produced directly at high-energy $ee$,\cite{PRODee} $ep$,\cite{PRODep}
and $pp$ colliders,\cite{PRODpp} whose subsequent decays can give rise
to  distinct like-sign dilepton signals.   If heavy Majorana neutrinos
are not  accessible at  high-energy  colliders, they  can still induce
lepton-flavour-violating decays   of  the $Z$ boson,\cite{KPS_Z,BSVMV}
the    Higgs   particle,\cite{APhiggs} and    the    $\tau$ and  $\mu$
leptons.\cite{IP,IP2} As we  will see, non-decoupling  quantum effects
due to potentially large SU(2)$_L$-breaking masses  play a key role in
these   flavour-changing-neutral-current   (FCNC)  phenomena.\cite{IP}
Heavy  Majorana neutrinos may    cause  breaking of  universality   in
leptonic  diagonal $Z$-boson\cite{BKPS}  and $\pi$  decays\cite{KP} or
influence\cite{SB&AS} the  size of the electroweak  oblique parameters
$S$, $T$ and  $U$.\cite{STU}   In fact, there  exist  many observables
summarized in Ref.\cite{LL} to which heavy Majorana neutrinos may have
sizeable contributions.  These observables include $\tau$-polarization
asymmetries, neutrino-counting experiments at  the CERN Large Electron
Positron Collider (LEP1) or at  the Stanford Linear Accelerator (SLC),
etc.

In the following we shall  show that high SU(2)$_L$-breaking masses in
a class of  neutrino models can lead to  large FCNC effects.   Because
these effects  are not correlated with  light neutrino masses, one can
overcome the see-saw suppression relations that usually accompany such
new-physics phenomena.\cite{Cheng/Li} Let us consider a two-generation
model of  the kind.   The model  is similar to  the  one  discussed in
Section 3. It  has two isosinglet neutrinos $\nu'_R$   and $S'_L$.  In
the weak basis $( (\nu_{\mu    L})^C,\ (\nu_{\tau L})^C,\   (S'_L)^C,\ 
\nu'_R )$ the neutrino mass matrix then takes the form
\begin{equation}
  \label{Mnmatr}
{\cal M}^\nu \ =\ \left( 
\begin{array}{cccc}  
0 & 0 & 0 & m_1\\
0 & 0 & 0 & m_2\\
0 & 0 & 0 & M\\
m_1 & m_2 & M & \mu
\end{array} \right). 
\end{equation}
Diagonalization of  ${\cal M}^\nu$ yields  two zero eigenvalues, which
would correspond to massless $\mu$ and $\tau$ neutrinos.  If $\mu \neq
0$ and $\mu/M  \ll  1$, they receive   small radiative masses  at high
orders.\cite{ZPCAP}  The  other two states  are very  heavy,  of order
$M\pm \mu$.  In  contrast to  the  usual seesaw scenario, the   ratios
$m_1/M$ and  $m_2/M$ remain  fully unconstrained.  Global  analyses of
low-energy data\cite{LL}  restrict  their  values to   $m_1/M,\  m_2/M
\stackrel{\displaystyle  <}{\sim} 0.1$.  For  reader's convenience, we
define the parameters
\begin{equation}
  \label{snul}
(s^{\nu_\mu}_L)^2\ \simeq\ \frac{m_1^2}{M^2}\ ,\qquad
(s^{\nu_\tau}_L)^2\ \simeq\ \frac{m_2^2}{M^2}\ ,
\end{equation}
The newly introduced parameters  quantify neutrino mixings between the
light and heavy Majorana states.  They also parameterize the deviation
of the  modified $Wl\nu_l$  coupling from  the  SM one.   An extensive
discussion is   given  in  Ref.\cite{LL,IP}  Including renormalization
effects into the definition of light--heavy neutrino mixings,\cite{KP}
one may tolerate the following upper limits
\begin{equation}
  \label{snubound}
(s^{\nu_\mu}_L)^2\ \ < \ \ 0.010, \quad
(s^{\nu_\tau}_L)^2\ \ < \ \ 0.035,\quad {\rm and}\quad
(s^{\nu_e}_L)^2 \ < \ \ 1\times 10^{-8}.
\end{equation}
The last  limit  comes from  the  requirement that   only the electron
family is responsible for  baryogenesis. Of course, electron  and muon
families may interchange their roles in Eq.\ (\ref{snubound}). 

\begin{figure}

\begin{center}
\begin{picture}(360,400)(0,0)
\SetWidth{0.8}

\ArrowLine(0,360)(20,360)\ArrowLine(60,360)(80,360)
\GCirc(40,360){20}{0.7}\Photon(40,340)(40,320){3}{2}
\Text(0,365)[b]{$l$}\Text(80,365)[b]{$l'$}
\Text(42,320)[l]{$Z,\gamma$}

\Text(100,360)[]{$=$}

\ArrowLine(120,360)(140,360)\ArrowLine(180,360)(200,360)
\ArrowLine(140,360)(180,360)\Text(160,367)[b]{$n_i$}
\DashArrowArc(160,360)(20,180,270){3}\PhotonArc(160,360)(20,270,360){2}{3}
\Text(145,340)[r]{$G^-$}\Text(175,340)[l]{$W^-$}
\Photon(160,340)(160,320){3}{2}
\Text(120,365)[b]{$l$}\Text(200,365)[b]{$l'$}
\Text(162,320)[l]{$Z,\gamma$}
\Text(160,300)[]{\bf (a)}

\ArrowLine(240,360)(260,360)\ArrowLine(300,360)(320,360)
\ArrowLine(260,360)(300,360)\Text(280,367)[b]{$n_i$}
\DashArrowArc(280,360)(20,270,360){3}\PhotonArc(280,360)(20,180,270){2}{3}
\Text(265,340)[r]{$W^-$}\Text(295,340)[l]{$G^-$}
\Photon(280,340)(280,320){3}{2}
\Text(240,365)[b]{$l$}\Text(320,365)[b]{$l'$}
\Text(282,320)[l]{$Z,\gamma$}
\Text(280,300)[]{\bf (b)}

\ArrowLine(0,260)(20,260)\ArrowLine(60,260)(80,260)
\ArrowLine(20,260)(60,260)\Text(40,267)[b]{$n_i$}
\PhotonArc(40,260)(20,180,270){2}{3}\PhotonArc(40,260)(20,270,360){2}{3}
\Text(25,240)[r]{$W^-$}\Text(55,240)[l]{$W^-$}
\Photon(40,240)(40,220){3}{2}
\Text(0,265)[b]{$l$}\Text(80,265)[b]{$l'$}
\Text(42,220)[l]{$Z,\gamma$}
\Text(40,200)[]{\bf (c)}

\ArrowLine(120,260)(140,260)\ArrowLine(180,260)(200,260)
\ArrowLine(140,260)(180,260)\Text(160,267)[b]{$n_i$}
\DashArrowArc(160,260)(20,180,270){3}\DashArrowArc(160,260)(20,270,360){2}
\Text(145,240)[r]{$G^-$}\Text(175,240)[l]{$G^-$}
\Photon(160,240)(160,220){3}{2}
\Text(120,265)[b]{$l$}\Text(200,265)[b]{$l'$}
\Text(162,220)[l]{$Z,\gamma$}
\Text(160,200)[]{\bf (d)}

\ArrowLine(240,260)(260,260)\ArrowLine(300,260)(320,260)
\Photon(260,260)(300,260){2}{4}\Text(280,267)[b]{$W^-$}
\ArrowLine(260,260)(280,240)\ArrowLine(280,240)(300,260)
\Text(270,240)[r]{$n_i$}\Text(295,240)[l]{$n_j$}
\Photon(280,240)(280,220){3}{2}
\Text(240,265)[b]{$l$}\Text(320,265)[b]{$l'$}
\Text(282,220)[l]{$Z$}
\Text(280,200)[]{\bf (e)}

\ArrowLine(0,160)(20,160)\ArrowLine(60,160)(80,160)
\DashArrowLine(20,160)(60,160){3}\Text(40,167)[b]{$G^-$}
\ArrowLine(20,160)(40,140)\ArrowLine(40,140)(60,160)
\Text(30,140)[r]{$n_i$}\Text(55,140)[l]{$n_j$}
\Photon(40,140)(40,120){3}{2}
\Text(0,165)[b]{$l$}\Text(80,165)[b]{$l'$}
\Text(42,120)[l]{$Z$}
\Text(40,100)[]{\bf (f)}

\ArrowLine(120,160)(135,160)\ArrowLine(135,160)(150,160)
\ArrowLine(150,160)(180,160)\ArrowLine(180,160)(200,160)
\Text(120,165)[b]{$l$}\Text(142,165)[b]{$l$}
\Text(165,165)[b]{$n_i$}\Text(200,165)[b]{$l'$}
\Photon(135,160)(135,120){3}{4}
\Text(137,120)[l]{$Z,\gamma$}
\PhotonArc(165,160)(15,180,360){2}{5}\Text(165,135)[]{$W^-$}
\Text(160,100)[]{\bf (g)}

\ArrowLine(240,160)(255,160)\ArrowLine(255,160)(270,160)
\ArrowLine(270,160)(300,160)\ArrowLine(300,160)(320,160)
\Text(240,165)[b]{$l$}\Text(262,165)[b]{$l$}
\Text(285,165)[b]{$n_i$}\Text(320,165)[b]{$l'$}
\Photon(255,160)(255,120){3}{4}
\Text(257,120)[l]{$Z,\gamma$}
\DashArrowArc(285,160)(15,180,360){3}\Text(285,135)[]{$G^-$}
\Text(280,100)[]{\bf (h)}

\ArrowLine(0,60)(20,60)\ArrowLine(20,60)(50,60)
\ArrowLine(50,60)(65,60)\ArrowLine(65,60)(80,60)
\Text(0,65)[b]{$l$}\Text(57,65)[b]{$l'$}
\Text(35,65)[b]{$n_i$}\Text(80,65)[b]{$l'$}
\Photon(65,60)(65,20){3}{4}
\Text(67,20)[l]{$Z,\gamma$}
\PhotonArc(35,60)(15,180,360){2}{5}\Text(35,35)[]{$W^-$}
\Text(40,0)[]{\bf (i)}

\ArrowLine(120,60)(140,60)\ArrowLine(140,60)(170,60)
\ArrowLine(170,60)(185,60)\ArrowLine(185,60)(200,60)
\Text(120,65)[b]{$l$}\Text(177,65)[b]{$l'$}
\Text(155,65)[b]{$n_i$}\Text(200,65)[b]{$l'$}
\Photon(185,60)(185,20){3}{4}
\Text(187,20)[l]{$Z,\gamma$}
\DashArrowArc(155,60)(15,180,360){3}\Text(155,35)[]{$G^-$}
\Text(160,0)[]{\bf (j)}

\end{picture}\\[0.7cm]
\end{center}

\fcaption{Feynman graphs pertaining to the decay $Z\to ll'$.  Graphs
  related to the effective $\gamma ll'$ vertex are also
  displayed.}\label{fig:8}

\end{figure}

\begin{figure}

\begin{center}
\begin{picture}(360,300)(0,0)
\SetWidth{0.8}

\ArrowLine(0,360)(20,360)\ArrowLine(60,360)(80,360)
\GCirc(40,360){20}{0.7}\Photon(40,340)(40,320){3}{2}
\Text(0,365)[b]{$\tau$}\Text(80,365)[b]{$l'$}
\Text(45,333)[l]{$Z,\gamma$}
\ArrowLine(40,320)(0,320)\ArrowLine(80,320)(40,320)
\Text(0,317)[t]{$l_1$}\Text(80,317)[t]{$l_1$}
\Text(40,295)[]{\bf (a)}

\ArrowLine(120,360)(140,360)\ArrowLine(180,360)(200,360)
\ArrowLine(140,360)(180,360)\Text(160,368)[b]{$n_i$}
\ArrowLine(140,320)(120,320)\ArrowLine(200,320)(180,320)
\ArrowLine(180,320)(140,320)\Text(160,314)[t]{$n_j$}
\Photon(140,360)(140,320){2}{4}\Photon(180,360)(180,320){2}{4}
\Text(120,365)[b]{$\tau$}\Text(200,365)[b]{$l'$}
\Text(120,317)[t]{$l_1$}\Text(200,317)[t]{$l_2$}
\Text(137,340)[r]{$W^-$}\Text(185,340)[l]{$W^+$}
\Text(160,295)[]{\bf (b)}

\ArrowLine(240,360)(260,360)\ArrowLine(300,360)(320,360)
\ArrowLine(260,360)(300,360)\Text(280,368)[b]{$n_i$}
\ArrowLine(260,320)(240,320)\ArrowLine(320,320)(300,320)
\ArrowLine(300,320)(260,320)\Text(280,314)[t]{$n_j$}
\DashArrowLine(260,360)(260,320){3}\Photon(300,360)(300,320){2}{4}
\Text(240,365)[b]{$\tau$}\Text(320,365)[b]{$l'$}
\Text(240,317)[t]{$l_1$}\Text(320,317)[t]{$l_2$}
\Text(257,340)[r]{$G^-$}\Text(305,340)[l]{$W^+$}
\Text(280,295)[]{\bf (c)}

\ArrowLine(0,260)(20,260)\ArrowLine(60,260)(80,260)
\ArrowLine(20,260)(60,260)\Text(40,268)[b]{$n_i$}
\ArrowLine(20,220)(0,220)\ArrowLine(80,220)(60,220)
\ArrowLine(60,220)(20,220)\Text(40,214)[t]{$n_j$}
\Photon(20,260)(20,220){2}{4}\DashArrowLine(60,260)(60,220){3}
\Text(0,265)[b]{$\tau$}\Text(80,265)[b]{$l'$}
\Text(0,217)[t]{$l_1$}\Text(80,217)[t]{$l_2$}
\Text(17,240)[r]{$W^-$}\Text(65,240)[l]{$G^+$}
\Text(40,195)[]{\bf (d)}

\ArrowLine(120,260)(140,260)\ArrowLine(180,260)(200,260)
\ArrowLine(140,260)(180,260)\Text(160,268)[b]{$n_i$}
\ArrowLine(140,220)(120,220)\ArrowLine(200,220)(180,220)
\ArrowLine(180,220)(140,220)\Text(160,214)[t]{$n_j$}
\DashArrowLine(140,260)(140,220){3}\DashArrowLine(180,260)(180,220){3}
\Text(120,265)[b]{$\tau$}\Text(200,265)[b]{$l'$}
\Text(120,217)[t]{$l_1$}\Text(200,217)[t]{$l_2$}
\Text(137,240)[r]{$G^-$}\Text(185,240)[l]{$G^+$}
\Text(160,195)[]{\bf (e)}

\ArrowLine(240,260)(260,260)\ArrowLine(320,260)(300,260)
\Line(260,260)(300,260)\Text(280,268)[b]{$n_i$}\Text(280,260)[]{{\boldmath
    $\times$}}
\ArrowLine(260,220)(240,220)\ArrowLine(300,220)(320,220)
\Line(300,220)(260,220)\Text(280,214)[t]{$n_j$}\Text(280,220)[]{{\boldmath
    $\times$}}
\Photon(260,260)(260,220){2}{4}\Photon(300,260)(300,220){2}{4}
\Text(240,265)[b]{$\tau$}\Text(320,265)[b]{$l_2$}
\Text(240,217)[t]{$l'$}\Text(320,217)[t]{$l_1$}
\Text(257,240)[r]{$W^-$}\Text(305,240)[l]{$W^+$}
\Text(280,195)[]{\bf (f)}

\ArrowLine(0,160)(20,160)\ArrowLine(80,160)(60,160)
\Line(20,160)(60,160)\Text(40,168)[b]{$n_i$}\Text(40,160)[]{{\boldmath
    $\times$}}
\ArrowLine(20,120)(0,120)\ArrowLine(60,120)(80,120)
\Line(60,120)(20,120)\Text(40,114)[t]{$n_j$}\Text(40,120)[]{{\boldmath
    $\times$}}
\Photon(20,160)(20,120){2}{4}\DashArrowLine(60,160)(60,120){3}
\Text(0,165)[b]{$\tau$}\Text(80,165)[b]{$l_2$}
\Text(0,117)[t]{$l'$}\Text(80,117)[t]{$l_1$}
\Text(17,140)[r]{$W^-$}\Text(65,140)[l]{$G^+$}
\Text(40,95)[]{\bf (g)}

\ArrowLine(120,160)(140,160)\ArrowLine(200,160)(180,160)
\Line(140,160)(180,160)\Text(160,168)[b]{$n_i$}\Text(160,160)[]{{\boldmath
    $\times$}}
\ArrowLine(140,120)(120,120)\ArrowLine(180,120)(200,120)
\Line(180,120)(140,120)\Text(160,114)[t]{$n_j$}\Text(160,120)[]{{\boldmath
    $\times$}}
\DashArrowLine(140,160)(140,120){3}\Photon(180,160)(180,120){2}{4}
\Text(120,165)[b]{$\tau$}\Text(200,165)[b]{$l_2$}
\Text(120,117)[t]{$l'$}\Text(200,117)[t]{$l_1$}
\Text(137,140)[r]{$G^-$}\Text(185,140)[l]{$W^+$}
\Text(160,95)[]{\bf (h)}

\ArrowLine(240,160)(260,160)\ArrowLine(320,160)(300,160)
\Line(260,160)(300,160)\Text(280,168)[b]{$n_i$}\Text(280,160)[]{{\boldmath
    $\times$}}
\ArrowLine(260,120)(240,120)\ArrowLine(300,120)(320,120)
\Line(300,120)(260,120)\Text(280,114)[t]{$n_j$}\Text(280,120)[]{{\boldmath
    $\times$}}
\DashArrowLine(260,160)(260,120){3}\DashArrowLine(300,160)(300,120){3}
\Text(240,165)[b]{$\tau$}\Text(320,165)[b]{$l_2$}
\Text(240,117)[t]{$l'$}\Text(320,117)[t]{$l_1$}
\Text(257,140)[r]{$G^-$}\Text(305,140)[l]{$G^+$}
\Text(280,95)[]{\bf (i)}

\Text(40,60)[]{\boldmath $+\quad ( l_1 \leftrightarrow l$\bf')}

\end{picture}\\
\end{center}

\fcaption{Feynman graphs pertaining to the decay $\tau\to
  l'l_1l_2$.}\label{fig:9}

\end{figure}                                  

Heavy Majorana  neutrinos can induce sizeable FCNC  decays of the type
$Z\to \mu\tau$,   $\tau\to \mu^- e^-e^+$  or $\tau\to \mu^-\mu^-\mu^+$
through quantum corrections    presented  in Figs.\  \ref{fig:8}   and
\ref{fig:9}.  Thus, the matrix element relevant  for the generic decay
$\tau (p_\tau)\to l(p_l)l_1(p_1)\bar{l}_2(p_2)$ acquires contributions
from $\gamma$- and $Z$-mediated graphs as well as from graphs with box
diagrams.  The respective transition elements are given by
\begin{eqnarray}
  \label{Tgamma}
i{\cal T}_\gamma (\tau\to l l_1 \bar{l}_2) &=&
     \frac{\alpha_w^2s_w^2}{4M_W^2}
     \delta_{l_1 l_2} \bar{u}_{l_1}\gamma^\mu v_{l_2}\
   \bar{u}_{l}\Big[ F^{\tau l}_\gamma (\gamma_\mu-\frac{q_\mu\not\! q}{q^2})
   (1-\gamma_5)\nonumber\\
&&              -iG_\gamma^{\tau l} \sigma_{\mu\nu}\frac{q^\nu}{q^2}
                (m_\tau(1+\gamma_5)+m_{l}(1-\gamma_5))\Big]u_\tau\, ,\\
  \label{TZ}
i{\cal T}_Z(\tau\to l l_1 \bar{l}_2)&=& \frac{\alpha_w^2}{16M_W^2}\
   \delta_{l_1 l_2}
   F_Z^{\tau l} \bar{u}_{l}\gamma_\mu(1-\gamma_5)u_\tau\ 
                                           \bar{u}_{l_1}\gamma^\mu
   (1-4s_w^2-\gamma_5)v_{l_2}\, ,\qquad\\
  \label{Tbox}
i{\cal T}_{\rm box}(\tau \to l l_1 \bar{l}_2) &=&
              \frac{\alpha_w^2}{16M_W^2}\  F_{\rm box}^{\tau ll_1l_2}\
   \bar{u}_{l}\gamma_\mu(1-\gamma_5)u_\tau\ \bar{u}_{l_1}\gamma^\mu
                                                    (1-\gamma_5)v_{l_2}\ ,
\end{eqnarray}
where $q=p_1+p_2$,    $s^2_w=1-M^2_W/M^2_Z$, and $F_\gamma^{\tau  l}$,
$G_\gamma^{\tau l}$,  $F_Z^{\tau l}$,  $F_{\rm box}^{\tau l  l_1 l_2}$
are  certain composite form  factors whose  analytic form  is given in
Ref.\cite{IP} Nevertheless,  it  is useful  to examine the  asymptotic
behaviour of  the   composite   form  factors for   large   values  of
$\lambda_{N_1}      =     m^2_{N_1}/M^2_W$    and     $\lambda_{N_2} =
m^2_{N_2}/M^2_W$ in  the  two-generation model  under  discussion. For
simplicity,  we   consider $\lambda_{N_1}   \sim  \lambda_{N_2}   \sim
\lambda_N \gg 1$. In this limit, we find
\begin{eqnarray}
  \label{CFgam}
F_\gamma^{\tau l} &\to & -\ \frac{1}{6}\; s_L^{\nu_\tau}s_L^{\nu_{l}}
                  \ln\lambda_N\, ,   \\
  \label{CGZ}
G_\gamma^{\tau l} &\to & \frac{1}{2}\; s_L^{\nu_\tau}s_L^{\nu_{l}}\, ,\\
    \label{CFZ}
F_Z^{\tau l} &\to & -\; \frac{3}{2}s_L^{\nu_\tau}s_L^{\nu_{l}}\ln\lambda_N\;
                -\; \frac{1}{2} s_L^{\nu_\tau}s_L^{\nu_{l}}\sum_{i=1}^{3}\
                             (s_L^{\nu_i})^2\lambda_N\, ,\\
    \label{CFbox}
F_{Box}^{\tau ll_1l_2} &\to &
               -\; (s_L^{\nu_\tau}s_L^{\nu_{l}}\delta_{l_1l_2}
                +s_L^{\nu_\tau}s_L^{\nu_{l_1}}\delta_{ll_2})\;
               +\; \frac{1}{2}s_L^{\nu_\tau}s_L^{\nu_{l}}s_L^{\nu_{l_1}}
                s_L^{\nu_{l_2}}\; \lambda_N\, .
\end{eqnarray}
If all light--heavy neutrino mixings $s_L^{\nu_l}$ are held fixed to a
constant, the one-loop  functions $F_\gamma^{\tau l}$, $G_\gamma^{\tau
  l}$,   $F_Z^{\tau  l}$, and    $F_{Box}^{\tau  ll_1l_2}$  in   Eqs.\ 
(\ref{CFgam})--(\ref{CFbox}) do  not  vanish  in   the  heavy neutrino
limit, and therefore   seem to violate  the decoupling  theorem due to
Appelquist and Carazzone.\cite{AC}    However, it is  known that   the
decoupling theorem does not apply to theories based on the spontaneous
or  dynamical   breaking  of    gauge   symmetries.  Since   we   hold
$s_L^{\nu_l}$  fixed   but  increase the    heavy-neutrino masses, the
violation   of    the    decoupling   theorem  originates     from the
SU(2)$_L$-breaking         Dirac    mass      terms        $m_1$   and
$m_2$.\cite{ZPCAP,APhiggs} The  expected decoupling of  the isosinglet
mass    $M$\cite{Senjan}    can    also     be     seen    in    Eqs.\ 
(\ref{CFgam})--(\ref{CFbox}).   This time  we   keep the Dirac  masses
$m_1$ and $m_2$ fixed and increase $M$.  Taking Eq.\ (\ref{snul}) into
account, it   is then easy to  show  that all  composite  form factors
vanish     for  large  values  of    $M\approx    m_{N_1},\ m_{N_2}$.  
Consequently, there is a competitive loop effect of two scales, namely
Dirac versus Majorana scale.  High Dirac masses lead to non-decoupling
loop effects while large Majorana masses give rise  to a screening and
reduce  the  strength of  the  effective FCNC coupling.  Nevertheless,
extensive analyses have  shown that a non-decoupling `window' confined
by the two mass scales exists, within which FCNC and other new-physics
phenomena come out to be rather large at a level that may be probed in
next-round experiments.\cite{APhiggs,IP,IP2}

As    examples, we  calculate   neutrinoless  tau-lepton   decays  and
flavour-violating   $Z$-boson   decays.       Taking the      dominant
non-decoupling parts of  the composite form  factors  into account, we
arrive at the simple expressions for the branching ratios
\begin{eqnarray}
  \label{Btauemu}
B(\tau^-\to \mu^- e^- e^+) &\simeq &
\frac{\alpha_w^4}{24576\pi^3}\ \frac{m_\tau^4}{M_W^4}\
 \frac{m_\tau}{\Gamma_\tau}  \Big[ \ |F_{\rm box}^{\tau \mu ee }|^2\nonumber\\
&&\hspace{-1.5cm} +\ 
2 (1-2s^2_w){\rm Re} [F_Z^{\tau \mu }F_{\rm box}^{\tau \mu ee *}]\
+\ 8s^4_w|F_Z^{\tau \mu}|^2  \ \Big] \nonumber\\
&&\hspace{-1.5cm}\simeq\      
\frac{\alpha_w^4}{98304\pi^3}\ \frac{m_\tau^4}{M_W^4}\
\frac{m_\tau}{\Gamma_\tau}
\frac{m^4_N}{M^4_W}\ (s_L^{\nu_\tau})^2 (s_L^{\nu_\mu})^2
\Big\{ (s_L^{\nu_{e}})^4\nonumber\\
&&\hspace{-1.5cm} +\      
2(1-2s^2_w)(s_L^{\nu_e})^2 \sum_i (s_L^{\nu_i})^2
+ 8s^4_w \Big[\sum_i (s_L^{\nu_i})^2\Big]^2\ \Big\}\, ,\\
  \label{Btaumumu}
B(\tau^-\rightarrow \mu^- \mu^- \mu^+) & \simeq &
     \frac{\alpha_w^4}{24576\pi^3}\ \frac{m_\tau^4}{M_W^4}\
     \frac{m_\tau}{\Gamma_\tau} \Big[ \frac{1}{2}|
             F_{\rm box}^{\tau \mu\mu\mu }|^2\nonumber\\
&&\hspace{-1.5cm} +\ 
2(1-2s^2_w){\rm Re} [F_Z^{\tau \mu}F_{\rm box}^{\tau\mu\mu\mu *}]\
 +\ 12 s^4_w|F_Z^{\tau \mu}|^2 \Big] \nonumber\\
&&\hspace{-1.5cm}\simeq\
\frac{\alpha_w^4}{98304\pi^3}\ \frac{m_\tau^4}{M_W^4}\
\frac{m_\tau}{\Gamma_\tau}
\frac{m^4_N}{M^4_W}\ (s_L^{\nu_\tau})^2 (s_L^{\nu_\mu})^2
\Big\{ \frac{1}{2}(s_L^{\nu_\mu})^4\nonumber\\
&&\hspace{-1.5cm} +\      
2(1-2s^2_w)(s_L^{\nu_\mu})^2  \sum_i (s_L^{\nu_i})^2
+ 12s^4_w \Big[\sum_i (s_L^{\nu_i})^2\Big]^2\ \Big\}\, ,\quad\\
  \label{BZtaumu}
B(Z\to \tau^- \mu^+  + \mu^- \tau^+) &=&
\frac{\alpha_w^3}{48\pi^2c_w^3}\frac{M_W}{\Gamma_Z}
                              |{\cal F}_Z^{\tau\mu}(M^2_Z)|^2 \nonumber\\
&&\hspace{-1.5cm}\simeq\
 \frac{\alpha_w^3}{768\pi^2c_w^3}\frac{M_W}{\Gamma_Z}
\frac{m^4_N}{M^4_W} (s_L^{\nu_\mu})^2 (s_L^{\nu_\tau})^2
\Big[ \sum_i (s_L^{\nu_i})^2 \Big]^2\, ,\quad
\end{eqnarray}
where $\Gamma_\tau=2.16\times 10^{-12}$  GeV and $\Gamma_Z = 2.49$ GeV
are respectively  the total widths  of the $\tau$  lepton and  the $Z$
boson known from experiment,\cite{PDG}  and ${\cal F}_Z^{\tau\mu}(0) =
F_Z^{\tau\mu}/2$.   The complete  analytic results  of  the  branching
ratios in  Eqs.\  (\ref{Btauemu})--(\ref{BZtaumu})  are  presented  in
Ref.\cite{IP}

To give an   estimate of the   size of the FCNC  effects,  we take the
maximally     allowed       values    $(s^{\nu_\tau}_L)^2=0.035$   and
$(s^{\nu_\mu}_L)^2=0.010$ given   by  Eq.\  (\ref{snubound}).    These
light-heavy neutrino mixings lead to the branching ratios
\begin{eqnarray}
  \label{BR}
B(\tau^- \to \mu^- \mu^- \mu^+) &\stackrel{\displaystyle <}{\sim}& 
                         2\times 10^{-6}\, ,\qquad 
B(\tau^- \to \mu^- e^- e^+)\ \ \stackrel{\displaystyle <}{\sim}\ \
                         1\times 10^{-6}\,,
\nonumber\\
B(Z\to \mu\tau) &\stackrel{\displaystyle <}{\sim}& 1.1\times 10^{-6}\ .
\end{eqnarray}
In  Eq.\ (\ref{BR}),  the  upper limits  are   estimated by  using the
heavy-neutrino  mass $m_N\simeq   4$  TeV,  which  results  from   the
requirement that  perturbative unitarity  be  valid.  The  theoretical
predictions   of  the branching ratios   must  be  contrasted with the
present experimental upper limits on these decays\cite{PDG}
\begin{eqnarray}
B(\tau^- \to \mu^- \mu^- \mu^+),\ B(\tau^- \to \mu^- e^- e^+) &<& 
1.4\times 10^{-5}\,,\nonumber\\
B(Z\to \mu\tau) &<& 1.3\times 10^{-5}\, ,
\end{eqnarray}
at the 90$\%$  confidence level.  Future high-luminosity colliders and
higher-precision experiments are  capable of improving the above upper
limits by one  order of  magnitude and  so  probe possible new-physics
effects due to heavy Majorana neutrinos.

\begin{figure}

\begin{center}
\begin{picture}(200,150)(0,0)
  \SetWidth{0.8}

\ArrowLine(50,80)(80,80)\Text(50,90)[l]{$e^-$}
\Line(80,80)(110,80)\Text(95,80)[]{{\boldmath $\times$}}
\Text(95,68)[]{$N_i$}
\ArrowLine(140,80)(110,80)\Text(125,90)[]{$e^-$} 
\Line(140,80)(170,80)\Text(155,80)[]{{\boldmath $\times$}}
\Text(155,92)[]{$N_j$}
\ArrowLine(170,80)(200,80)\Text(200,90)[r]{$e^-$}
\DashArrowArcn(110,80)(30,180,360){5}
\DashArrowArc(140,80)(30,180,270){5}
\DashArrowArc(140,80)(30,270,360){5}
\Photon(140,50)(140,20){3}{4}\Text(145,20)[l]{$\gamma$}
\Text(110,115)[b]{$\chi^-$}
\Text(118,60)[tr]{$\chi^-$}\Text(163,60)[lt]{$\chi^-$}

\end{picture}\\[0.5cm]
\end{center}

\fcaption{Typical two-loop diagram contributing to the EDM 
of electron.}\label{fig:10}

\end{figure}

\newpage

\subsection{Electric dipole moment of the electron}
\noindent
In  general, CP-violating new-physics interactions  may give rise to a
large contribution to the  EDM of the   electron.  This results in  an
interaction in the Lagrangian of the form
\begin{equation}
\label{EDM}
{\cal L}_{d}\ =\ ie\, \Big(\frac{d_e}{e}\Big)\, \bar{e}\, \sigma_{\mu\nu}
\gamma_5\, e\,  \partial^\mu A^\nu\, .
\end{equation}
The experimental upper bound on electron EDM  is very strict: $(d_e/e)
< 10^{-26}$  cm.\cite{PDG}  This   bound  is very  crucial,  as  heavy
Majorana  neutrinos   can induce an    EDM   of the  electron  at  two
loops.\cite{Ng2} A typical diagram is  shown in Fig.\ \ref{fig:10}.  A
simple estimate of   this  contribution based on  a  naive dimensional
analysis for $m_{N_2},\ m_{N_1}\gg M_W$ gives\cite{APRD}
\begin{equation}
\label{EDMaj}
\frac{d_e}{e} \sim (10^{-24}\, \mbox{cm})\, \times\, 
{\rm Im}(h_{1e}h^*_{2e})^2\, 
\frac{m_{N_1}m_{N_2}(m^2_{N_1}-m^2_{N_2})}{(m^2_{N_1} + m^2_{N_2})^2}\, 
\ln\Big(\frac{m_{N_1}}{M_W}\Big)
\, .
\end{equation}
In Eq.\ (\ref{EDMaj}) the factor depending on $m_{N_i}$ is always much
smaller than  unity.  Clearly, the above EDM  limit could be important
for  $|h_{li}| \gg  0.1$ and/or  ultra-heavy  Majorana neutrinos  with
$m_{N_i}>10^{11}$ TeV.  In this prediction,   one should bear in  mind
that stability of   the Higgs  potential  under radiative  corrections
requires  $|h_{li}|={\cal   O}(1)$.\cite{HLP} Nevertheless,    the EDM
contribution  is several  orders of  magnitude below the  experimental
bound  for    $x_N   <    10^{-3}$  and/or     $|h_{l1}|,\    |h_{l2}|
\stackrel{\displaystyle <}{\sim}   10^{-2}$.  In this  context,  it is
interesting to notice  that leptogenesis models with nearly degenerate
heavy Majorana neutrinos can naturally evade possible EDM constraints.

\setcounter{section}{10}
\setcounter{equation}{0}
\section{Conclusions}\label{sec:10}
\noindent 
We have reviewed many recent  developments that have been taking place
in the scenario  of  baryogenesis through leptogenesis, and  discussed
the implications that heavy Majorana neutrinos may have for laboratory
experiments.    In   the  standard   leptogenesis   scenario,\cite{FY}
$L$-violating  decays of heavy Majorana  neutrinos,  which are out  of
thermal equilibrium, produce  an excess in $L$  that is converted into
the observed  BAU  through  $B+L$-violating interactions   mediated by
sphalerons.    We  have paid  more  attention  to   different kinds of
mechanisms of CP violation involved in  the $N_i$ decays.  One has two
generic   types: (a) CP   violation originates  from the  interference
between a tree-level graph  and  the absorptive  part of the  one-loop
vertex ($\varepsilon'$-type  CP violation) and  (b) CP violation comes
from the interference between  a  tree-level graph and  the absorptive
part of    the  one-loop  $N_i$   self-energy  ($\varepsilon$-type  CP
violation).

Recently, there has   been renewed interest in $\varepsilon$-type   CP
violation  in   models with mixed  heavy  Majorana  neutrinos.  If the
masses of two  heavy  neutrinos become degenerate, then   finite-order
perturbation theory  does  no longer apply,   and various methods have
been      invoked   to  cope      with      this problem    in     the
literature.\cite{Paschos}  Here,  we have   discussed the whole  issue
based on   an  effective resummation   approach  to  unstable particle
mixing.\cite{APRD} One then finds that $\varepsilon$-type CP violation
is  resonantly  enhanced if the  mass splitting  of the heavy Majorana
neutrinos  is comparable to  their  widths (cf.\ Eq.\ (\ref{CPcond})),
and if  the parameter $\delta_{CP}$ defined  in Eq.\ (\ref{dCP}) has a
value close to 1.  These two conditions turn out to be necessary and
sufficient for resonant  CP violation of  order unity.\cite{APRD} As a
consequence, the  scale  of leptogenesis  may be   lowered  up to  TeV
energies.  In  fact, E$_6$-motivated scenarios with  nearly degenerate
heavy Majorana neutrinos of order 1 TeV and universal Yukawa couplings
can still be responsible for the  BAU.  This last observation receives
firm support  after  solving numerically the BE's.\cite{APRD}  In this
kinematic  range, the  $\varepsilon'$-type contributions are extremely
suppressed.   Also, finite-temperature  effects  on  masses of  heavy
neutrinos and on  decay widths do  not spoil the above  conditions for
resonant CP asymmetries.  Finally, constraints due to electron EDM are
still too weak to play a role in leptogenesis.

The   fact that the  isosinglet-neutrino  scale can be  lowered to TeV
energies has a  number of virtues.  If one  has  to appeal to  (local)
supersymmetry in order to    maintain  the flatness of  the   inflaton
potential, one then  has to worry  about the cosmological consequences
of the  gravitino during the  nucleosynthesis epoch.  Since the weakly
interacting gravitinos are  at most of  the order of  a few TeV, their
slow decay rate will  lead to an overproduction of  D and $^3{\rm He}$
unless the number density-to-entropy ratio   at the time of  reheating
after inflation is less than about $10^{-10}$.\cite{EWK&SW} This leads
to     quite      low    reheating     temperatures    $T_{\rm     RH}
\stackrel{\displaystyle  <}{\sim}    10^{10}$  GeV,  after  which  the
radiation-dominated era starts and baryogenesis or leptogenesis can in
principle  occur.  The  latter causes a  major  problem for  GUT-scale
baryogenesis and GUT's as well, especially  if $T_{\rm RH} \sim M_{\rm
  GUT}$.    In  this context,   it     is important to    remark  that
supersymmetric  extensions of models  with isosinglet neutrinos in the
multi-TeV range as the  ones discussed here  can comfortably evade the
known gravitino problem mentioned above.  In such scenarios, the heavy
inflatons with a mass of order $10^{12}$ GeV  can now decay abundantly
to    the  relatively    lighter heavy  Majorana    neutrinos yielding
equilibrated  (incoherent)  thermal    distributions  for the   latter
particles.

Heavy Majorana neutrinos may also induce sizeable FCNC effects such as
$Z\to \tau\mu$ and  $\tau\to \mu\mu\mu$ at a  level that can be probed
in  near-future experiments.  Non-decoupling   loop  effects of   high
SU(2)$_L$-breaking   masses  play a   significant  role  in increasing
drastically      the     strength        of  these         new-physics
phenomena.\cite{ZPCAP,APhiggs,IP,IP2}  However, these  heavy  Majorana
neutrinos cannot account for the  BAU at the  same time.  Depending on
the   model,   they can  coexist   with  the  heavy Majorana neutrinos
responsible for  leptogenesis without destroying baryogenesis.  At the
LHC,\cite{PRODpp} the viability of such models  can be tested directly
by looking for  like-sign  dilepton signals.   Such signals will  then
strongly   point   towards   the  scenario   of   baryogenesis through
leptogenesis as   the   underlying mechanism  for   understanding  the
baryonic asymmetry in nature.

\nonumsection{Acknowledgements}
\noindent  
The author gratefully acknowledges    discussions with Bill   Bardeen,
Zurab  Berezhiani,  James   Bjorken,  Francisco   Botella,    Wilfried
Buchm\"uller, Stan    Brodsky, Darwin  Chang,    Sacha Davidson,   Ara
Ioannisian,  Pasha Kabir,   Wai-Yee Keung,  Emmanuel Paschos,  Michael
Peskin, Georg   Raffelt,   Utpal Sarkar,   Mikhail   Shaposhnikov, Leo
Stodolsky,  Arkady Vainshtein, and Xinmin    Zhang.  The author   also
wishes  to thank Jose  Bernab\'eu,  Amitava Datta,  Kai Diener, Zoltan
Gagyi-Palffy,    Monoranjan  Guchait,   Amon   Ilakovac, Bernd Kniehl,
J\"urgen K\"orner, Marek Nowakowski,  Joannis Papavassiliou,  and Karl
Schilcher for collaboration.

\nonumsection{References}

\newpage

\appendix

\noindent 
We shall     now list useful      analytic expressions  for   one-loop
self-energies of the Higgs and fermion  fields as well as for one-loop
vertex couplings $\chi^+ lN_i$,  $\chi^0\nu_l  N_i$ and $H\nu_l  N_i$.
We present  relations  between wave-function  CT's and  unrenormalized
self-energies in the OS  renormalization scheme.  The analytic results
are expressed in  terms   of  standard loop  integrals   presented  in
Ref.\cite{HV} Instead,   we adopt  the signature for   the Minkowskian
metric $g_{\mu\nu} = \mbox{diag}(1,-1,-1,-1)$.

The Feynman rules pertaining to the minimal model may be read off from
the Lagrangian   (\ref{LYint}).    We first   give  the    the   Higgs
self-energies $\chi^-\chi^-$,  $\chi^0\chi^0$ and $HH$, shown in Fig.\
\ref{fig:1}(d)--(f)
\begin{eqnarray}
\label{PiHiggs}
\Pi_{\chi^-\chi^-}(p^2)& = & \Pi_{\chi^0\chi^0}(p^2)\ =\ 
\Pi_{HH}(p^2)\nonumber\\
&=& \sum\limits_{l=1}^{n_L} \sum\limits_{i=1}^{n_R}\, 
\frac{|h_{li}|^2}{8\pi^2}
\, \Big[\, m^2_{N_i} B_0(p^2,m^2_{N_i},0)\, +\, p^2\, B_1(p^2,m^2_{N_i},0)\,
\Big]\, .\quad
\end{eqnarray}
{}From Eq.\ (\ref{PiHiggs}), the  universality of the divergent  parts
of the  wave  functions $\delta  Z_{\chi^-}$, $\delta  Z_{\chi^0}$ and
$\delta  Z_{H}$    is evident, since   $\delta  Z^{div}_\Phi   = -{\rm
  Re}\Pi'^{div}_\Phi   (0)$  for all  field components    of the Higgs
doublet $\Phi$.

{}Figures \ref{fig:1}(g), (h) and (j) show the individual contributions
to the one-loop fermionic transitions,  $l'\to l$, $\nu_{l'}\to \nu_l$
and   $N_j\to N_i$,    respectively.   Explicit calculation  of  these
self-energy transitions gives
\begin{eqnarray}
\label{Sigml'l}
\Sigma_{ll'}(\not\! p) &=& - \sum\limits_{i=1}^{n_R}\,
\frac{h_{li}h^*_{l'i}}{16\pi^2} \not\! p\, P_L\, B_1(p^2,m^2_{N_i},0)\, ,\\
\label{Sigmn'n}
\Sigma_{\nu_l\nu_{l'}}(\not\! p) &=& - \sum\limits_{i=1}^{n_R}\,
\frac{1}{32\pi^2}\, \Big[ (h^*_{li}h_{l'i}\not\! p P_R\, +\,
h_{li}h^*_{l'i}\not\! p P_L)\Big(B_1(p^2,m^2_{N_i},0)\, \nonumber\\
&&+\, B_1(p^2,m^2_{N_i},M^2_H)\Big)\, +\, m_{N_i}
(h^*_{li}h^*_{l'i} P_R\, +\, h_{li}h_{l'i} P_L)\nonumber\\
&&\times\, \Big(B_1(p^2,m^2_{N_i},0)\, -\, 
B_1(p^2,m^2_{N_i},M^2_H)\Big)\Big],\\
\label{Sigmji}
\Sigma_{N_iN_j}(\not\! p) &=& - \sum\limits_{l=1}^{n_L}\,
\frac{1}{16\pi^2}\, (h^*_{li}h_{lj}\not\! p P_R\, +\,
h_{li}h^*_{lj}\not\! p P_L)\Big(\, \frac{3}{2} B_1(p^2,0,0)\nonumber\\
&& +\, \frac{1}{2} B_1 (p^2,0,M^2_H)\Big) .
\end{eqnarray}
Note that   the light-neutrino self-energies  in  Eq.\ (\ref{Sigmn'n})
contain non-zero masses in the  limit $\not\!  p\to 0$ if $M_H\not=0$.
Therefore,    at $T=0$,  small   radiative    neutrino masses can   be
generated.\cite{ZPCAP}  However, these contributions are suppressed by
small Yukawa  coupling,   and   therefore   do  not   invalidate   any
experimental or cosmological limit.

Before  we express the  wave-function renormalization factors in terms
of unrenormalized self-energies,\cite{KP}  we  first  notice that  the
one-loop  fermionic transitions $f_j\to   f_i$ (with $f_i=l,\  \nu_l,\
N_i$) given by Eqs.\  (\ref{Sigml'l})--(\ref{Sigmji}) have the generic
form
\begin{equation}
\label{Sigma}
\Sigma_{ij}(\not\! p)\ =\ \Sigma^L_{ij} (p^2)\not\! p P_L\, +\,
\Sigma^R_{ij} (p^2)\not\! p P_R\, +\, \Sigma^M_{ij}(p^2) P_L\, +\,
\Sigma^{M*}_{ji}(p^2) P_R\, ,
\end{equation}
where only   dispersive parts are    considered.   If the  transitions
involve Majorana fermions only, one then has the additional properties
$\Sigma^L_{ij}  (p^2)= \Sigma^{R*}_{ij} (p^2)$ and $\Sigma^M_{ij}(p^2)
= \Sigma^M_{ji}(p^2)$. Following Ref.\cite{KP}, the wave-function CT's
are given by
\begin{eqnarray}
\label{dZfii}
\delta Z^f_{ii} &=& -\Sigma^L_{ii}(m^2_i)\, -\, 2m^2_i\Sigma^L_{ii}{}'(m^2_i)
\, -\, 
m_i\Big[ \Sigma^M_{ii}{}'(m^2_i)+\Sigma^{M*}_{ii}{}'(m^2_i)\Big]\nonumber\\
&&+\, \frac{1}{2m_i}\Big[ \Sigma^M_{ii}(m^2_i)-\Sigma^{M*}_{ii}(m^2_i)\Big],
\end{eqnarray}
and, for $i\not= j$,
\begin{equation}
\label{dZfij}
\delta Z^f_{ij}\ =\ \frac{2}{m^2_i-m^2_j}\Big[
m^2_j\Sigma^L_{ij}(m^2_j)\, +\, m_im_j\Sigma^{L*}_{ij}(m^2_j)\, +\,
m_i\Sigma^M_{ij}(m^2_j)\, +\, m_j\Sigma_{ij}^{M*}(m^2_j)\Big].
\end{equation}
The wave-function renormalization of  charged leptons may  be obtained
by Eqs.\ (\ref{dZfii}) and  (\ref{dZfij}),  if all terms  depending on
$\Sigma^M_{ij}(p^2)$ and its  derivative are neglected. At this point,
it is important to remark that  there will be additional contributions
to our wave-function CT's of the  Higgs and fermion fields coming from
loops related to  SM gauge particles.  As we  have seen in Section  8,
the same kind of loops can induce non-zero thermal masses to the Higgs
and  fermion  particles.\cite{HAW,MEC}   These new  contributions  are
universal and therefore pose  no    problem to the     Yukawa-coupling
renormalization discussed in Section 4.

Finally, one-loop corrections to   the vertices $\chi^\pm l^\mp  N_i$,
$\chi^0  \nu_l      N_i$  and  $H       \nu_lN_i$   shown  in   Figs.\ 
\ref{fig:1}(a)--(c) have been calculated in Ref.\cite{APRD} by keeping
the  complete functional dependence  on  the Higgs-boson mass  $M_H$. 
Their analytic expressions are given by
\begin{eqnarray}
\label{chi_N}
i{\cal V}_{\chi^+l^-N_i} &=& -i \bar{u}_l P_R u_{N_i}\, 
\sum\limits_{l'=1}^{n_L}\sum\limits_{j=1}^{n_R}\,
\frac{m_{N_i}m_{N_j}}{16\pi^2}\ h^*_{l'i}h_{l'j}h_{lj} 
\nonumber\\
&&\times\, \Big[ C_0(0,0,m^2_{N_i},0,m^2_{N_j},0)\, +\, 
C_{12}(0,0,m^2_{N_i},0,m^2_{N_j},0)\, \Big] ,\\
\label{chi0N}
{\cal V}_{\chi^0\nu_l N_i} &=& -\, \bar{u}_l P_R u_{N_i}\, 
\sum\limits_{l'=1}^{n_L}\sum\limits_{j=1}^{n_R}\,
\Big\{\, \frac{m_{N_i}m_{N_j}}{32\sqrt{2}\pi^2}\
h^*_{l'i}h_{l'j}h_{lj} \nonumber\\
&&\times\, \Big[C_0(0,0,m^2_{N_i},0,m^2_{N_j},0)\, +\, 
C_0(0,0,m^2_{N_i},M^2_H,m^2_{N_j},0)\nonumber\\
&&+C_{12}(0,0,m^2_{N_i},0,m^2_{N_j},0)\, +\, 
C_{12}(0,0,m^2_{N_i},M^2_H,m^2_{N_j},0)\Big]\nonumber\\
&&+\frac{1}{32\sqrt{2}\pi^2}\ h_{l'i}h^*_{l'j}h_{lj}
\nonumber\\
&&\times \Big[\, m^2_{N_i}\Big(C_{12}(0,0,m^2_{N_i},0,m^2_{N_j},0)\, -\, 
C_{12}(0,0,m^2_{N_i},M^2_H,m^2_{N_j},0)\Big)\nonumber\\
&&-\, M^2_H C_0 (0,0,m^2_{N_i},M^2_H,m^2_{N_j},0) \Big]\, \Big\}, \\
\label{HN}
-i{\cal V}_{H\nu_lN_i} &=& i\bar{u}_l P_R u_{N_i}\,  
\sum\limits_{l'=1}^{n_L}\sum\limits_{j=1}^{n_R}\,
\Big\{\, \frac{m_{N_i}m_{N_j}}{32\sqrt{2}\pi^2}\ h^*_{l'i}h_{l'j}h_{lj}
\nonumber\\
&&\times\, \Big[C_0(0,0,m^2_{N_i},0,m^2_{N_j},0)\, +\, 
C_0(0,0,m^2_{N_i},M^2_H,m^2_{N_j},0)\nonumber\\
&&+C_{12}(0,0,m^2_{N_i},0,m^2_{N_j},0)\, +\, 
C_{12}(0,0,m^2_{N_i},M^2_H,m^2_{N_j},0)\Big]\nonumber\\
&&+\frac{1}{32\sqrt{2}\pi^2}\ h_{l'i}h^*_{l'j}h_{lj}
\nonumber\\
&&\times \Big[\, m^2_{N_i}\Big(C_{12}(0,0,m^2_{N_i},0,m^2_{N_j},0)\, -\, 
C_{12}(0,0,m^2_{N_i},M^2_H,m^2_{N_j},0)\Big)\nonumber\\
&&-\, M^2_H C_0 (0,0,m^2_{N_i},M^2_H,m^2_{N_j},0) \Big]\, \Big\} .
\end{eqnarray}
{}From Eqs.\ (\ref{chi0N})  and (\ref{HN}) it is  easy to see that the
$L$-conserving part   of the couplings    $\chi^0 \nu_l N_i$   and  $H
\nu_lN_i$ is UV  finite and vanishes  identically for $M_H\to 0$.

\end{document}